\documentclass[aps,prb,floatfix,twocolumn,superscriptaddress,showpacs,amsmath,amssymb]{revtex4-1}
\usepackage{graphicx}
\usepackage{epstopdf}
\usepackage{epsfig} 
\usepackage{url}
\usepackage[USenglish]{babel}
\allowdisplaybreaks

\usepackage[colorlinks=true,linkcolor=blue,citecolor=red]{hyperref}

\newcommand{\ca}{c^{\phantom{\dagger}}}
\newcommand{\cc}{c^\dagger}
\newcommand{\da}{d^{\phantom{\dagger}}}
\newcommand{\dc}{d^\dagger}
\newcommand{\be}{\begin{equation}}
\newcommand{\ee}{\end{equation}}
\newcommand{\bea}{\begin{eqnarray}}
\newcommand{\eea}{\end{eqnarray}}
\newcommand{\ba}{\begin{eqnarray*}}
\newcommand{\ea}{\end{eqnarray*}}
\newcommand{\dagga}{{\phantom{\dagger}}}
\newcommand{\bR}{\mathbf{R}}

\newcommand{\bk}{\mathbf{k}}

\newcommand{\kp}{k_{\perp}}

\newcommand{\br}{\mathbf{r}}

\newcommand{\dis}{\displaystyle}

\newcommand{\up}{\uparrow}
\newcommand{\down}{\downarrow}
\newcommand{\fract}[2]{\frac{\dis #1}{\dis #2}}
\newcommand{\Tr}{\mathrm{Tr}}
\newcommand{\eqn}[1]{(\ref{#1})}
\newcommand{\ket}[1]{|{#1}\rangle}

\newcommand{\braket}[3]{\langle{#1}| {#2} |{#3} \rangle}

\newcommand{\quave}[1]{\langle {#1} \rangle}
\newcommand{\Hcal}{\mathcal{H}}
\newcommand{\nonu}{\nonumber}

\def\eg{\mbox{\it e.g.}}  
\def\ie{\mbox{\it i.e.}}
\def\vhyb{v_\text{\tiny hyb}}

\begin{document}

\title{Electronic transport and dynamics in correlated heterostructures.}

\author{G.~Mazza}
\affiliation{Scuola Internazionale Superiore di Studi Avanzati (SISSA), 
Via Bonomea 265, 34136 Trieste, Italy}

\author{A.~Amaricci}
\affiliation{Democritos National Simulation Center, 
Consiglio Nazionale delle Ricerche, 
Istituto Officina dei Materiali (IOM) and 
Scuola Internazionale Superiore di Studi Avanzati (SISSA),  
Via Bonomea 265, 34136 Trieste, Italy}

\author{M.~Capone}
\affiliation{Democritos National Simulation Center, 
Consiglio Nazionale delle Ricerche, 
Istituto Officina dei Materiali (IOM) and 
Scuola Internazionale Superiore di Studi Avanzati (SISSA), 
Via Bonomea 265, 34136 Trieste, Italy}

\author{M.~Fabrizio}
\affiliation{Scuola Internazionale Superiore di Studi Avanzati (SISSA), 
Via Bonomea 265, 34136 Trieste, Italy}

\pacs{}

\begin{abstract}
We investigate by means of the time-dependent Gutzwiller approximation
the transport properties of a strongly-correlated slab subject to
Hubbard repulsion and  connected with to two metallic leads kept at a different  electrochemical potential. 
We focus on the real-time evolution of the electronic properties after
the slab is connected to the leads and consider both metallic and Mott
insulating slabs. When the correlated slab is metallic, the system relaxes to a
steady-state that sustains a finite current. 
The zero-bias conductance is finite and independent of the degree of
correlations within the slab as long as the system remains metallic. 
On the other hand, when the slab is in a Mott insulating state, the
external bias leads to currents  that are exponentially activated  by charge tunneling across the
Mott-Hubbard gap, consistent with the Landau-Zener dielectric
breakdown scenario. 
\end{abstract}
\maketitle

Correlated materials such as the transition-metal oxides (TMOs)
feature an impressive variety of interesting properties, usually caused by presence of
electrons in the partially filled outer $d$-orbitals of the
transition-metal atoms.\cite{Bednorz86,Anderson97,Hosono08} 
The electrons in these orbitals give rise to narrow electronic bands,
which increase the relevance of electron-electron interactions with respect to inner orbital
shells. 
The competition between the tendency of the
electrons to localise near the ionic position to minimize the
potential energy and the energy gained by
delocalising through the lattice is at the heart of the diverse and
remarkable features of these materials. 
The most paradigmatic effect of the strong correlation in
the bulk of TMOs is the Mott metal-insulator transition\cite{MottRMP,Imada1998RMP}: by
changing pressure, temperature or chemical doping a metallic state can
be transformed into a partially filled insulating state. 

The effects of the strong correlation are nevertheless not limited to 
bulk properties, and they can induce subtle and remarkable effects at
the surface or at the interface of materials.\cite{OkamotoPRB2004,IshidaPRB2008,Biscaras2010NC} 
Lately, the quest for a theoretical understanding of how bulk correlations
influence the reconstruction of the surface electronic phase triggered
a great deal of attention.\cite{Zubko,Heterostructure,Sulpizio} 
This is not only motivated by the advances in the engineering and
control of heterostructures with potential applications ranging from
electronics to sustainable energy, but it also helps to reconcile
contrasting experimental evidences.\cite{Freericks2004} 
A paradigmatic example in this sense is provided in the metallic state
of the prototypical correlated compound V$_2$O$_3$, 
where surface-sensitive photoemission measurements fail to observe
quasiparticle excitations, which are instead  observed in
bulk-sensitive experiments.\cite{Rodolakis_exp_dead_layer} 
This evidence was theoretically interpreted in
Ref.~\onlinecite{Borghi_prl09}, where it has been shown that for an
inhomogeneous correlated system the
metallic character of the surface electronic states gets strongly
suppressed with respect to the bulk.

More recently the development of time-resolved experiments triggered a huge
interest into the non-equilibrium phenomena occurring in
correlated systems.\cite{Orenstein2012PT,Guiot2013NC,neqDMFT_RMP}  
In particular, the possibility to follow in real time the evolution of
the electronic response offered a new opportunity to understand the
formation and the properties of non-linearities in  correlated heterostructures.  
This is a necessary step to to improve the design of
electronic devices for technological applications. 
Nonetheless,  the difficulty in the theoretical treatment of system
breaking  both space and time translation \cite{neqDMFT_RMP} invariance has slowed down the advance in this field.
The initial steps focused mainly on stationary
states in heterostructures, with the aim
to identify the mechanism underlying the formation or the suppression of conductive
channels in the presence of a sufficiently large potential bias.\cite{SOkamoto_PRB,SOkamoto_PRL,Heary2009,gabi_arxiv,amaricci_scbias} 
The early stages of the  investigation of non-equilibrium dynamics of strongly
correlated systems focused on the real-time evolution of driven homogeneous systems. 
In this context important results were obtained using
non-equilibrium formulation of dynamical mean-field
theory~\cite{neqDMFT_RMP}  to investigate, {\eg}, the non-linear response to constant
\cite{Joura2008,amaricci_stationary} or periodic fields \cite{Tsuji2008PRB,Tsuji2011PRL} or
to address the dielectric breakdown of Mott insulators.\cite{eckstein_dielectric_breakdown} 
Insight into the electronic dynamics of  inhomogeneous
systems out of equilibrium has been obtained by mean of Time-Dependent Gutzwiller (TDG) 
method.\cite{SchiroFabrizio_short} The initial focus was
on the quench dynamics of a layered system of correlated 
planes coupled to phonons.\cite{patrice_andre} 
The extension of non-equilibrium DMFT to  the inhomogenoeus case allowed to
study in more details the real-time dynamics of driven
heterostructures either in presence of a voltage potential bias
\cite{Eckstein_slab} and after shining ultra-short light
pulses.\cite{Eckstein_prl2014} 

In this work we study the non-equilibrium dynamics of a strongly
correlated heterostructure coupled to external metallic leads and
driven out of equilibrium  by a voltage potential bias.  
Using a suitable formulation of the TDG \cite{SchiroFabrizio_short} method we study the dynamics
of the inhomogeneous system and its non-linear transport properties.
In the first part of this work we focus on the correlated metallic
regime where $U$ is smaller than the critical value for the Mott
transition $U_c$. Here we follow the dynamical formation of surface states with enhanced
metallic character after the sudden coupling to external metallic leads. 
We show that this effect is associated to  a characteristic time scale which
diverges at the Mott transition.  
Next, we show that the formation of current-carrying
stationary states in presence of a finite voltage bias depends
directly on the value of the coupling between the slab and the leads. While for small
couplings a stationary state can always be reached, at strong
coupling the system gets trapped in a metastable state caused by an
effective decoupling of the slab from the leads.    
We study the current-voltage characteristic of the
system and demonstrate both the existence of a universal behavior with
respect to interaction at small bias and the presence of a negative
differential resistivity for larger applied bias. 

In the second part of this paper we focus on the Mott insulating
regime for $U > U_c$. Following the same analysis of the metallic case, we study
first the dynamical formation of a metallic surface state in the 
Mott insulating regime. Indeed, we show that this is determined by 
an avalanche effect leading to an exponential growth of the quasiparticle
weight inside the slab bulk. 
Such quasiparticle weight becomes exponentially small in the  bulk
over a distance of the order of the Mott transition correlation lenght.\cite{Borghi_prl09}
Finally, we show that for large enough voltage bias a
conductive stationary state can be created from a Mott insulating
slab with an highly non-linear current-bias characteristics.
In particular, we show that the currents are exponentially activated
with the applied bias and associate this behavior to a Landau-Zener 
dielectric breakdown mechanism.\cite{oka_prl2003,oka_prl2005}

The rest of the paper is divided as follows: In section
\ref{sec:model_method} we introduce the inhomogeneous 
formulation of the TDG method  and briefly discuss the derivation
of some important relations. The technical aspects of this derivation and
the details of the numerical solutions are outlined
in appendix \ref{app:eqs_of_motion}. 
In Sec.~\ref{sec:noneq_metal} we apply the TDG method to study the
non-equilibrium electronic transport in biased metallic inhomogeneous
systems. We discuss first on the zero-bias regime and we relate it to
the equilibrium description of the same system. Then we study the
transport in, respectively, the small- and large-bias regimes. 
In section \ref{sec:noneq_insulator} we present our results for the
case of a driven Mott insulating slab and discuss the properties of
insulating dielectric breakdown caused by the applied voltage bias. 
Finally, in Sec.~\ref{sec:conclusions} we summarize our results and
discuss future perspectives.

\section{Model and Method}
\label{sec:model_method}
We consider a strongly correlated slab composed by a series of $N$ two-dimensional layers 
with in-plane and inter-plane hopping amplitudes and a purely local interaction term.
We indicate the layer index with  $z=1,\ldots,N$ while 
we assume discrete translational symmetry on the $xy$ plane of each layer. This
enables us to introduce a two-dimensional momentum $\bk$ so that
the slab hamiltonian reads
\begin{eqnarray}
\nonumber
H_{\text{Slab}} &=& \sum_{z=1}^N \sum_{\bk,\sigma} \epsilon_{\bk} \dc_{\bk,z,\sigma} \da_{\bk,z,\sigma} \\
\nonumber
 &+& \sum_{z=1}^{N-1} \sum_{\bk,\sigma} \Big( t_{z,z+1}\,  \dc_{\bk,z+1,\sigma} \da_{\bk,z,\sigma} + H.c. \Big) \\
 &+& \sum_{z=1}^N  \sum_{\br} \, \bigg(\frac{U}{2} \,\left(n_{\br,z} - 1 \right)^2 +  E_{z}\,  n_{\br,z}
 \bigg),
 \label{eq:Hslab}
\end{eqnarray}
where $\epsilon_{\bk}=-2t\,\big(\cos k_x +\cos k_y \big)$  is the electronic dispersion for 
nearest-neighbor tight-binding Hamiltonian on a square lattice, $\br$
label the sites on each two-dimensional layer, $t_{z,z+1}$ is the
inter-layer hopping parameter and $E_{z}$ is a layer-dependent on-site
energy.
In the rest of this work we assume $t_{z,z+1} = t$ and we use $t =1$
as our energy unit.

\begin{figure}
  \begin{center}
    \includegraphics[width=0.9\linewidth]{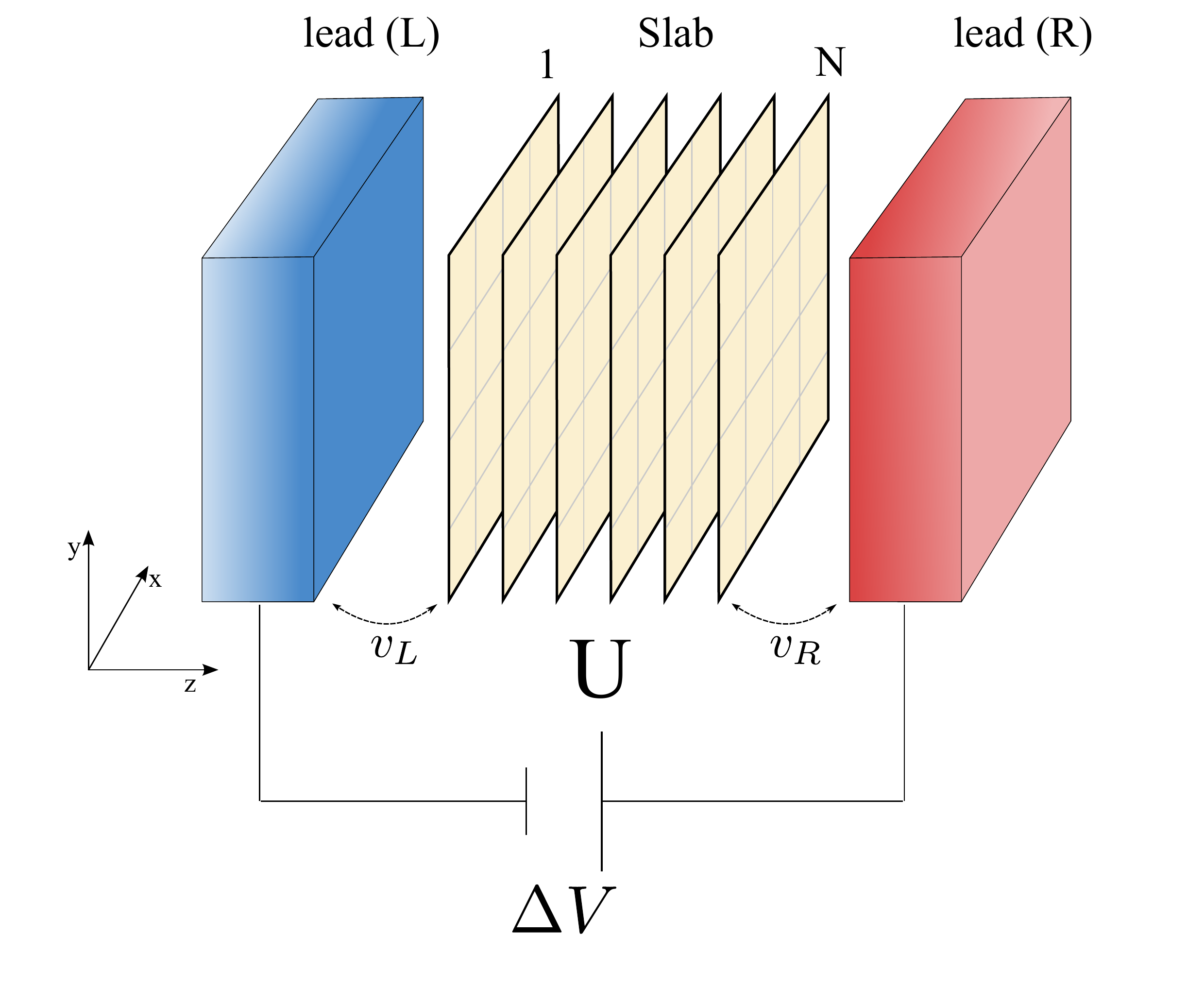}
  \end{center}
  \caption{
    (Color online) Sketch of the correlated slab sandwiched between 
    semi-infinite metallic leads. $v_L$ and $v_R$ represent respecitvely
    left and right slab-leads hybridization coupling. 
  }
  \label{fig:structure}
\end{figure}

A finite bias $\Delta V$ across the system is applied by coupling with  an external environment
composed by two, left ($L$)  and right ($R$), semi-infinite metallic
leads described by not interacting Hamiltonians
with symmetrically shifted energy bands
\begin{equation}
  H_{\text{Lead}} = \sum_{\alpha=L,R} \;\sum_{\bk, \kp ,\sigma} \big(
  \varepsilon^{\alpha}_{\bk} + t^{\alpha}_{\kp} - \mu_{\alpha} \big)
  \, 
  \cc_{\bk \kp \alpha \sigma} \ca_{\bk \kp \alpha \sigma}, 
  \label{eq:h_leads}
\end{equation}
where $k_\perp$ labels the $z-$component of the electron momentum. In Eq. \eqn{eq:h_leads}
$\varepsilon^{\alpha}_{\bk} = -2 t_\alpha\big(\cos k_x + \cos k_y\big)$, 
 $t^{\alpha}_{\kp} = -2 t_\alpha\,\cos k_\perp$, where we shall assume $t_L=t_R=t$, and 
$\mu_{L/R}=\pm e \Delta V/2$, with $e$ the electron charge.
We couple the system to the metallic leads through a finite tunneling amplitude 
between the left(right) lead and the first(last) layer, i.e. 
\begin{equation}
  H_{\text{Hyb}} = \sum_{\alpha=L,R} \sum_{\bk ,\kp, \sigma}
  \Big( v^{\alpha}_{\kp} \,\cc_{\bk \kp \alpha  \sigma} \da_{\bk z_{\alpha} \sigma }  + H.c.\Big),
  \label{eq:h_hyb}
\end{equation}
where $z_L=1$, $z_R=N$ and 
\begin{equation}
  v^{\alpha}_{\kp} = \sqrt{\frac{2}{N_{\perp}}}\;  \sin\kp\, v_{\alpha},
  \label{eq:vhyb_def}
\end{equation}
which corresponds to open boundary conditions 
for the leads along the $z-$direction. 

The final Hamiltonian is thus the sum of Eqs.
\eqn{eq:Hslab}, \eqn{eq:h_leads} and \eqn{eq:h_hyb}
\begin{equation}
  H = H_{\text{Slab}} + H_{\text{Leads}} + H_{\text{Hyb}}.
  \label{eq:H}
\end{equation}

We drive the system out-of-equilibrium by suddenly switching  the tunneling  between 
the slab and the leads, that is $v_L(t)= v_R(t)=v_\text{hyb}\, \theta(t)$,
and by turning on a finite bias $\Delta V(t) = \Delta V \, r(t)$ according to  
a time-dependent protocol $r(t)$ that, if not explicitly stated,
we also take as a step function.
We exploit the local energies $E_z$ in Eq.~(\ref{eq:Hslab})
to model the potential drop between left and right leads.
Even though the profile of the inner potential should be self-consistently determined
by the long range coulomb interaction, see e.g. Refs.~\onlinecite{charlebois_pn_junction} 
and \onlinecite{freericks_charge_reconstruction},
we assume that a flat profile $E_z=0$ represents a reasonable choice for the system
in its metallic phase, simulating the screening of the electric field
inside the metal.
On the other hand, in the insulating phase we shall assume a linear potential
drop $E_z= e \Delta V  (N+1-2z)/2(N+1)$ matching the left and right leads 
chemical potential for $z=0$ and $z=N+1$.
In the rest of the work we will assume the units $e=1$ and $\hbar=1$.

Since an exact solution of the time-dependent Schr\"odinger equation
for the model (\ref{eq:H}) is not feasible we resort the so called
time-dependent Gutzwiller approximation~\cite{SchiroFabrizio_short}
and its extension to inhomogeneous systems.~\cite{patrice_andre} 
While we refer the reader to Ref.~\onlinecite{Fabrizio_review} for a
detailed derivation, we sketch  the main steps that lead to
the Gutzwiller dynamical equations  for the present case of an
inhomogeneous system coupled to semi-infinite leads.

As customary we split the Hamiltonian (\ref{eq:H}),
$H=\Hcal_0+\Hcal_{loc}$, into a not-interacting term $\mathcal{H}_0$
and a purely local interaction part $\mathcal{H}_{loc}$ 
\begin{equation}
  \begin{split}
  \mathcal{H}_0 
  &= H_{\text{leads}} + H_{\text{hyb}} + 
  \sum_{z=1}^N \sum_{\bk,\sigma} \epsilon_{\bk}\, \dc_{\bk,z,\sigma} \da_{\bk,z,\sigma}  \\
  &\phantom{=} + \sum_{z=1}^{N-1} \sum_{\bk,\sigma} \Big(t_{z,z+1}\,
  \dc_{\bk,z+1,\sigma} \da_{\bk,z,\sigma} + H.c.\Big), 
  \end{split}
\end{equation}
\begin{equation}
  \mathcal{H}_{loc} =  \sum_{z=1}^N  \sum_{\br}\, \frac{U}{2}
  \big(n_{\br,z} - 1 \big)^2 +  E_{z}  n_{\br,z}\equiv \sum_{\bR}\,
  \mathcal{H}_{loc,\bR},
\end{equation}
where $\bR=(\br,z)$, 
and define the time-dependent variational wavefunction 
\begin{equation}
  \ket{\Psi(t)} = \prod_\bR \mathcal{P}_\bR(t) \ket{\Psi_0(t)},
  \label{eq:gz_variational_ansatz}
\end{equation}
where $\ket{\Psi_0(t)}$ is a time-dependent wavefunction for which
Wick theorem holds, and $\mathcal{P}_\bR$ are linear operators that
act on the local Hilbert space at site  $\bR$ and control, through  a
set of time-dependent variational parameters, the weights of the local
electronic configurations.
The  dynamics of the variational parameters and 
of the wavefunction $\ket{\Psi_0(t)}$  is obtained by applying the
time-dependent variational  principle $\delta \mathcal{S}=0$ on the
action $\mathcal{S} = \int \, \braket{\Psi}{\,i\partial_t -
  H\,}{\Psi}$. Upon imposing the following constraints 

\begin{equation}
  \begin{split}
    &\braket{\Psi_0(t)}{\,\mathcal{P}^{\dag}_\bR (t) \mathcal{P}_\bR(t) \,}{\Psi_0(t)} = 1,\\
    &\braket{\Psi_0(t)}{\,\mathcal{P}^{\dag}_\bR (t)
      \mathcal{P}_\bR(t)\, \dc_{\bR \sigma} \da_{\bR
        \sigma'}\,}{\Psi_0(t)} \\
    & \qquad\qquad \qquad = 
    \braket{\Psi_0(t)}{\,\dc_{\bR \sigma} \da_{\bR \sigma'}\,}{\Psi_0(t)},
  \end{split}
  \label{eq:constraints_generic}
\end{equation}
expectation values can be analytically computed in lattices with
infinite coordination number.~\cite{Fabrizio_review} In particular,
if one parametrizes the Gutzwiller operators  $\mathcal{P}_\bR$
through site- and time-dependent variational matrices
$\hat{\Phi}_\bR(t)$ in the basis of the local electronic
configurations, the following  closed set of coupled dynamical
equations are readily obtained~\cite{Fabrizio_review}
\begin{eqnarray}
  i \,\frac{\partial \mid \Psi_0(t) \rangle}{\partial t} &=&
  \mathcal{H}_*[\hat{\Phi}(t)] \mid \Psi_0(t) \rangle 
 , \label{eq:dot_psi}\\
  i \frac{\partial \hat{\Phi}_\bR (t)}{\partial t} &=& 
  \mathcal{H}_{loc,\bR} \hat{\Phi}_\bR (t) \nonumber\\
  && + 
  \langle \Psi_0(t) \mid \frac{\partial
    \mathcal{H}_*[\hat{\Phi}(t)]}{\partial \hat{\Phi}_\bR^{\dag}(t)}
  \mid \Psi_0(t) \rangle.
  \label{eq:dot_phi}
\end{eqnarray}
In Eqs. \eqn{eq:dot_psi} and \eqn{eq:dot_phi}
$\mathcal{H_*}[\hat{\Phi}(t)]$ is an effective not-interacting
Hamiltonian that depends parametrically on the variational matrices
$\hat{\Phi}_\bR(t)$. Eq.~(\ref{eq:dot_psi}) represents an effective  
Schr\"odinger equation for non-interacting electrons and is 
commonly interpreted as a Hamiltonian for the
coherent quasiparticles.
The dynamics of the local variational parameters
determined by Eq.~(\ref{eq:dot_phi}) can be associated
to the incoherent excitations of the Hubbard bands. 
The two dynamical evolutions are coupled in a mean-field like fashion,
each degree of freedom  providing a time-dependent field for the other one. 
This aspect, although being an approximation that does not allow to
reproduce a genuine relaxation to a steady state,  still represents a
great advantage of the present method with respect to the standard
time-dependent Hartree-Fock. 

If we use as local basis at site $\bR$ the empty state $\ket{0}$, the
doubly-occupied one, $\ket{2}$,  and the singly-occupied ones,
$\ket{\sigma}$ with $\sigma=\up,\down$ referring to the electron spin,
and discard magnetism and $s$-wave superconductivity, the
matrix $\hat{\Phi}_\bR(t)$ can be chosen  diagonal with matrix
elements $\Phi_{\bR,00}(t) \equiv \Phi_{\bR,0}(t)$, $\Phi_{\bR,22}(t)
\equiv \Phi_{\bR,2}(t)$, and 
$\Phi_{\bR,\up\up}(t) = \Phi_{\bR,\down\down}(t)\equiv \Phi_{\bR,1}(t)/\sqrt{2}$.
Due to translational invariance within each $xy$ plane,  
the variational matrices depend explicitly only on the layer index
$z$, 
i.e. $\hat{\Phi}_\bR(t) = \hat{\Phi}_z(t)$, 
and the constraints (\ref{eq:constraints_generic}) are satisfied by
imposing
\begin{equation}
  \left| \Phi_{z,0}(t) \right|^2 + \left| \Phi_{z,2}(t) \right|^2 +
  \left| \Phi_{z,1}(t) \right|^2 = 1, 
\end{equation}
and
\begin{equation}
  \begin{split}
  \delta_z(t) &\equiv   \left| \Phi_{z,0}(t) \right|^2 - \left|
    \Phi_{z,2}(t) \right|^2  \\
  &=   1-\sum_{\bk \sigma}\, \braket{\Psi_0(t)}{\,\dc_{\bk z \sigma}
    \da_{\bk z \sigma}\,}{\Psi_0(t)},
  \end{split}
\end{equation}
where $\delta_z(t)$ is the instantaneous doping of layer $z$. 
Through this choice we obtain the following effective Hamiltonian
$\mathcal{H}_*[\hat{\Phi}(t)]$
\begin{equation}
  \begin{split}
  \mathcal{H}_*[\hat{\Phi}(t)] &= H_{\text{Leads}} +  \sum_{z=1}^N
  \sum_{\bk,\sigma} \left| R_z(t) \right|^2  \epsilon_{\bk}\,
  \dc_{\bk,z,\sigma} \da_{\bk,z,\sigma} \\
  &+ \sum_{z=1}^{N-1} \sum_{\bk,\sigma}\,\Big( R^*_{z+1}(t)\,R_z(t)\;
  \dc_{\bk,z+1,\sigma} \da_{\bk,z,\sigma} + H.c.\Big) \\
  &+ \sum_{\alpha=L,R}\, \sum_{\bk, \kp, \sigma}
  \Big(v_{\kp}\, R_{z_{\alpha}}(t)\; \cc_{\bk \kp \alpha  \sigma}
  \da_{\bk z_{\alpha} \sigma }  +H.c.\Big),
  \label{eq:Hstar}
  \end{split}
\end{equation}
where the layer-dependent hopping renormalization factor reads
\begin{equation}
  R_z(t) = \sqrt{\fract{2}{1-\delta_z(t)^2}}\; \bigg( \Phi_{z,0}(t)
  \Phi^*_{z,1}(t) + \Phi_{z,2}^*(t) \Phi_{z,1}(t) \bigg).
\end{equation}

Straightforward differentiation of Eq.~(\ref{eq:Hstar})
with respect to $\hat{\Phi}^{\dag}_z(t)$ yields the
equations of motion for the variational matrices 
(\ref{eq:dot_phi}), which together with the 
effective Schr\"odinger equation~(\ref{eq:dot_psi})
completely determine the variational dynamics
within the TDG approximation.
Though the derivation of the set of coupled dynamical
equations is very simple, the final result is cumbersome so that  
we present it in the appendix \ref{app:eqs_of_motion}
together with details on its numerical integration.

We characterize the non equilibrium behavior of the system by studying the 
electronic transport through the slab. In particular, in the following we shall  
define the electronic current flowing from the left/right lead to the 
first/last layer of the slab as the {\sl contact current} with expression 
\begin{equation}
  j_{\alpha}(t) = -i \left[ \sum_{\bk \sigma} \sum_{\kp} v_{\kp} 
    \braket{\Psi(t)}{\,\dc_{\bk z \sigma} \ca_{\bk \kp \alpha \sigma}\,}{\Psi(t)} - c.c.  \right],
\end{equation}
and the {\sl layer current} as the current flowing from 
the $z$-th to the $z+1$-th layer, i.e. 
\begin{equation}
  j_z(t) = -i \left[\sum_{\bk \sigma} 
  \braket{\Psi(t)}{\dc_{\bk z \sigma} \da_{\bk z+1 \sigma}}{\Psi(t)} - c.c.  \right].
\end{equation}
Within the TDG approximation
these two observables read, respectively,
\begin{equation}
  j_{\alpha}(t) = -i \bigg[R_{z_{\alpha}}^*(t) \sum_{\bk \sigma}
  \sum_{\kp} v_{\kp} \quave{\dc_{\bk z \sigma} \ca_{\bk \kp \alpha
      \sigma}} - c.c. \bigg]
  \label{eq:j_contact}
\end{equation}
and
\begin{equation}
  j_z(t) = -i \bigg[R_z^*(t) R_{z+1}(t) \sum_{\bk \sigma}
  \quave{\dc_{\bk z \sigma} \da_{\bk z+1 \sigma}} - c.c.  \bigg],
  \label{eq:j_layer}
\end{equation}
where $\quave{\ldots} \equiv \braket{\Psi_0(t)}{\ldots}{\Psi_0(t)}$. 
Notice that, due to the left/right symmetry, $j_L = -j_R$ and  we need
only to consider currents for $z \leq N/2 $.


\section{Non-equilibrium transport in the strongly correlated metal}
\label{sec:noneq_metal}
In this section we consider the case of a correlated slab in its metallic
phase, $U\!<\!U_c$ where $U_c\!\approx\!16$ is the critical value above 
which the system is Mott insulating in equilibrium. 

\subsection{Zero-bias dynamics}
To start with we shall consider the dynamics at zero-bias $\Delta V\!=\!0$. 
In this case we assume that the non-equilibrium perturbation is the
sudden switch of the tunnel amplitude $\vhyb$ between the correlated
slab and the leads. 
In the equilibrium regime, the metallic character at the un-contacted surfaces is
strongly suppressed with respect to the bulk as effect of the reduced
kinetic energy. This suppression, commonly described in terms of a surface 
{\it dead layer}, extends over a distance which is quite remarkably 
controlled by a critical correlation
length $\xi$ associated to the Mott transition. Indeed $\xi$ is found to grow
approaching the metal-insulator transition and diverges at the
transition point.~\cite{Borghi_prl09} 
In presence of a contact with external metallic leads the surface
state is characterized by a larger quasiparticle weight with respect
to that of the bulk irrespective of its metallic or insulating
character, realizing what is called a {\it living layer}.~\cite{Borghi_prb10} 
As we shall see in the following, by switching on $\vhyb$ it should be
possible  to turn the dead layer into the living one on a characteristic time scale
$\tau$: The dynamical counterpart of the correlation length $\xi$.

\begin{figure}
  \begin{center}
    \includegraphics[scale=0.6]{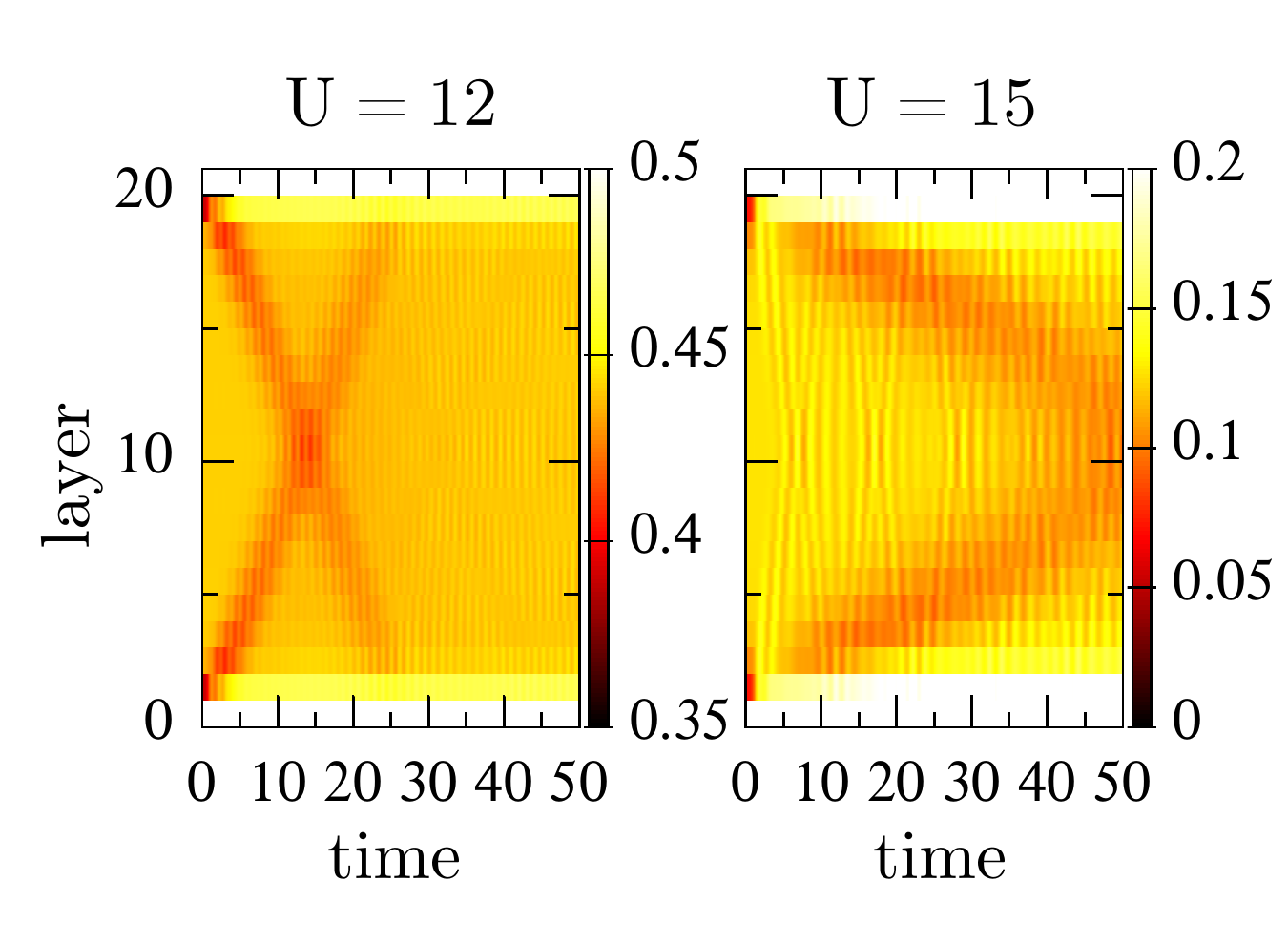}
    \caption{(Color online) Layer-resolved dynamics of the local
      quasiparticle weights $|R_z(t)|^2$  for
      a slab of $N=20$ layers and two values of the interaction $U$.
      The slab-lead hybridization is equal to the inter-layer hopping
      amplitude $\vhyb=1.0$.}\label{fig:wake_up_3d}
  \end{center}
\end{figure}
In Fig.~\ref{fig:wake_up_3d} we show the time-evolution of the layer-dependent quasiparticle 
weight $Z_z(t)\!\equiv\!\left| R_z(t)\right|^2$ for a $N\!=\!20$ slab and 
different values of the interaction $U$. 
The dynamics shows a characteristic light-cone effect, \ie~a constant
velocity propagation of the perturbation from the junctions at the
external layers  $z\!=\!1$ and $z\!=\!N$ to the center of the slab.
After few reflections the light-cone disappears leaving the system in a stationary state. 
The velocity of the propagation is found to be proportional to the
bulk quasiparticle weight hence it decreases as the Mott transition is
approached for $U\!\rightarrow\!U_c$.

\begin{figure}[t]
  \begin{center}
    \includegraphics[scale=0.6]{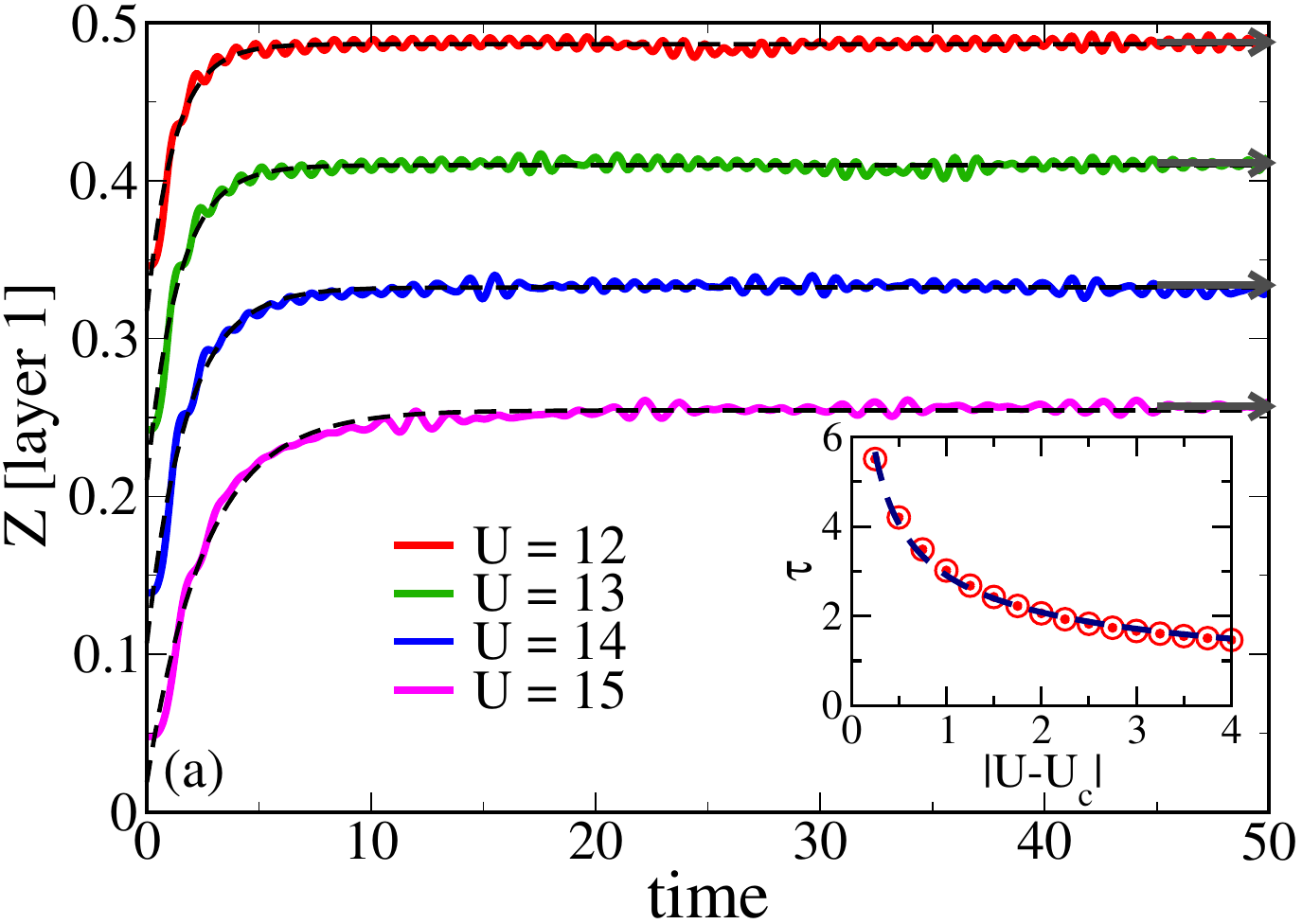}
    \includegraphics[scale=0.6]{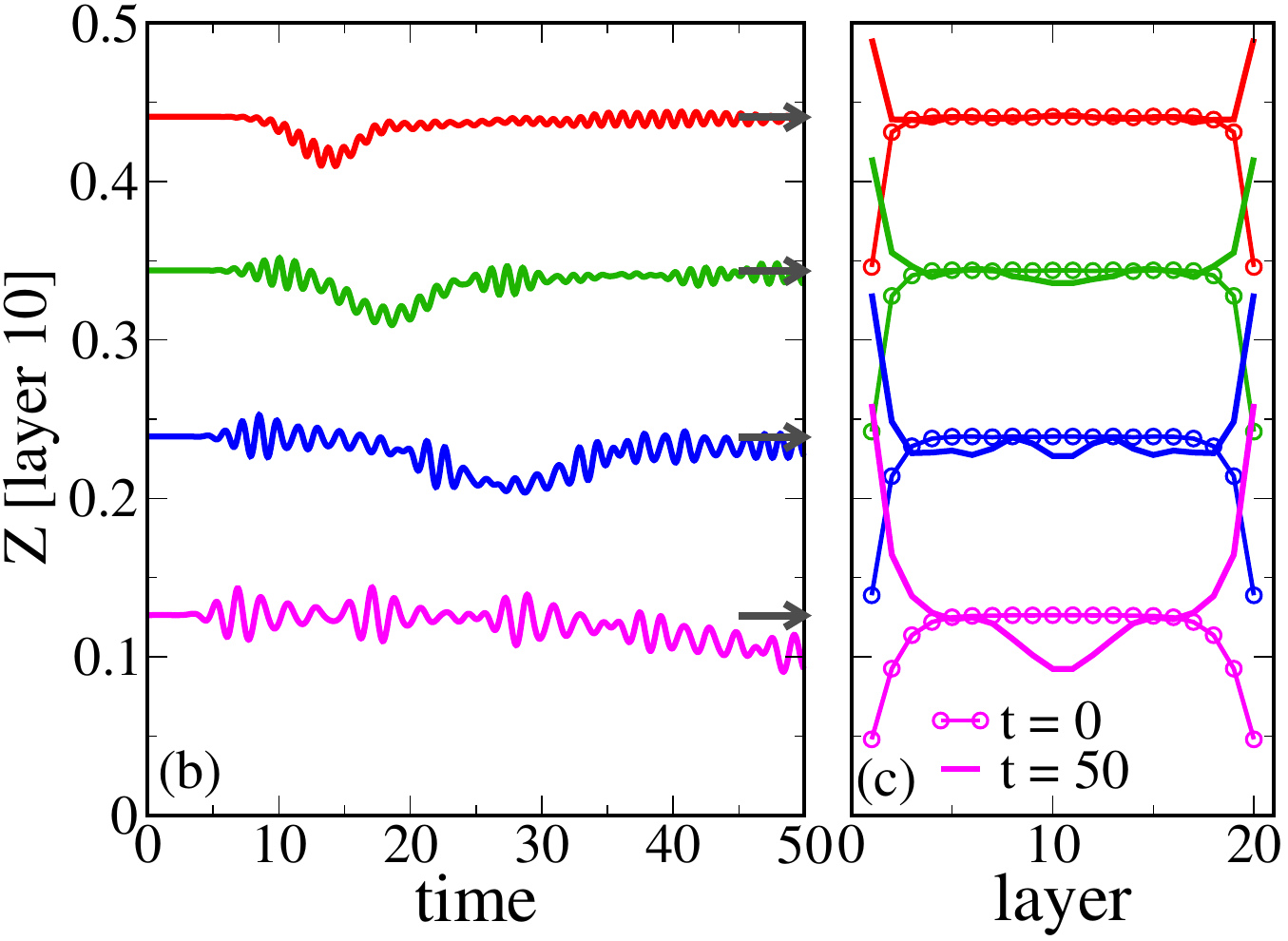}
    \caption{(Color online)
      (a) Dynamics of the local quasiparticle weight 
      for the first layer. Dashed lines are the fitting curves obtained with 
      Eq.~(\ref{eq:z_wake}). Arrows represent the hybridized slab equilibrium values.
      Inset: dead layer awakening time as a function
      of $U$. Dashed lines represents the fitting curve $\tau = \alpha/|U-U_c|^{\nu_*}$
      with $\nu_* \approx 0.4895$
      (b) Dynamics of the local quasiparticle weight for the bulk ($z=10$) layer.
      Arrows represent the hybridized slab equilibrium values. 
      (c) Quasiparticle weight profiles at times $t=0$ (dotted lines) and $t=50$
      (lines).
    }
    \label{fig:wake_up_metal}
  \end{center}
\end{figure}

The boundary layers are strongly perturbed by the sudden switch of the tunneling amplitude. 
In particular, we observe in Fig.~\ref{fig:wake_up_metal}(a) that the surface
{\sl dead layers} rapidly transform into {\sl living layers} with
stationary quasiparticle weights greater than the bulk ones and
equal to the equilibrium values for the same set-up.~\cite{Borghi_prb10}
This has to be expected since the energy injected is not extensive. 
On the contrary, the bulk layers are weakly affected by the coupling with 
the metal leads, see Fig.~\ref{fig:wake_up_metal}(b). Their dynamics 
is only affected by small oscillations and temporary deviations 
from the stationary values due to the perturbation propagation
described by the light-cone reflections.

We characterize the evolution from the {\sl dead} to the 
{\sl living} layer by fitting the dynamics of the boundary layer
quasiparticle weight with an exponential relaxation towards
a stationary value:
\begin{equation}
  Z(t) = Z_{\text{dead}} + \Big(Z_{\text{living}}-Z_{\text{dead}}\Big)\,\Big(1-e^{-t/\tau}\Big).
  \label{eq:z_wake}
\end{equation}
As illustrated in Fig.~\ref{fig:wake_up_metal}(b) the dynamics shows a
slowing-down upon approaching the Mott transition.
In particular the dead-layer wake-up time $\tau$ diverges as
$\tau\!\sim\!|U\!-\!U_c|^{-\nu_*}$ when we approach the critical value $U_c$
with a critical  exponenent that we estimate as $\nu_*\!=\!0.4895$, very close to the mean-field value $\nu_*\!=\!1/2$. 
Such a mean-field dependence, similar to that of the correlation lenght
$\xi\!\sim\! |U\!-\!U_c|^{-1/2}$ ~\cite{Borghi_prb10} implies, through 
$\tau\sim \xi^{\,\zeta}$, a dynamical critical exponent $\zeta\!=\!1$.

\subsection{Small-bias regime}
\begin{figure}[tbh]
  \begin{center}
    \includegraphics[scale=0.6]{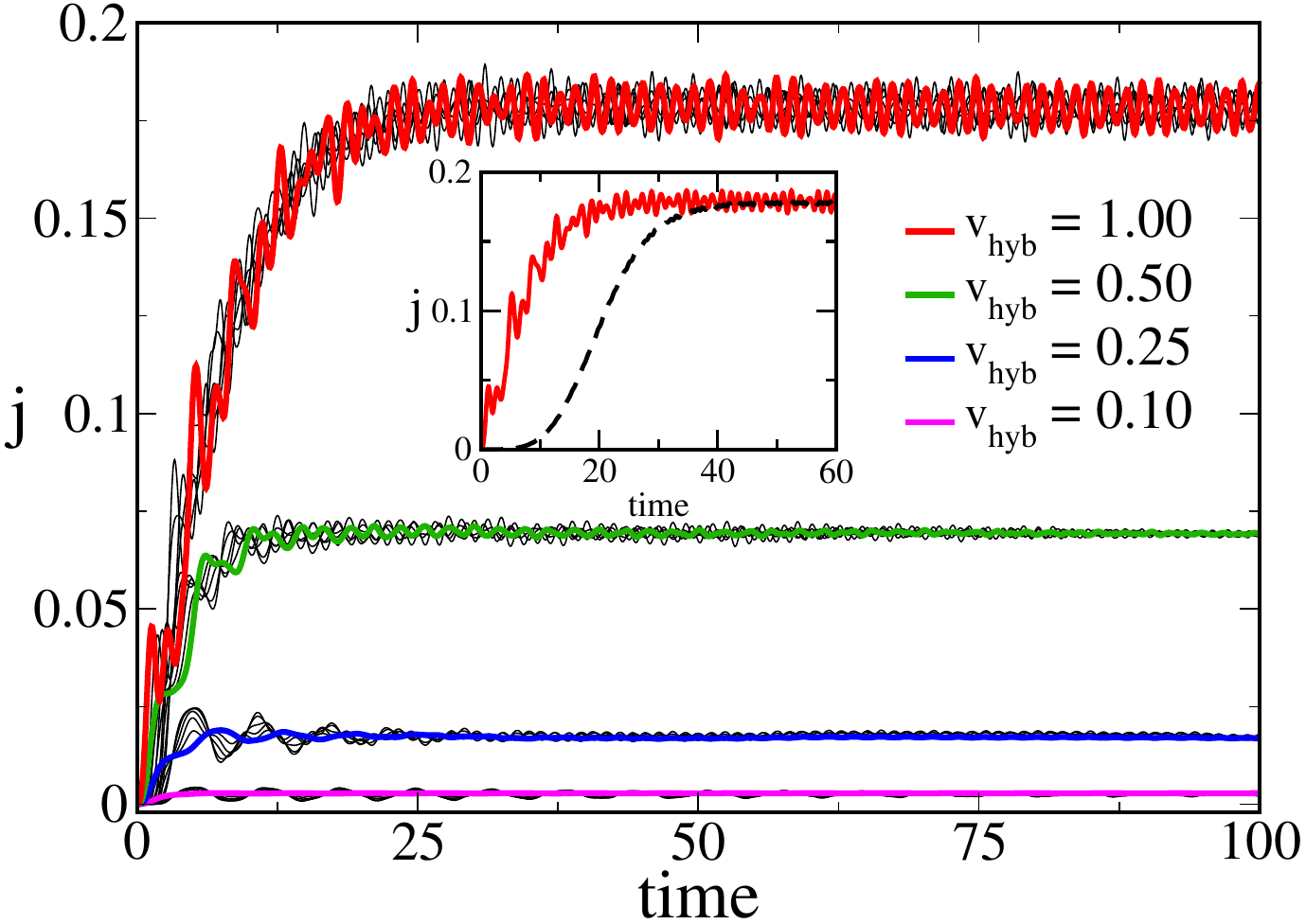}
    \includegraphics[scale=0.6]{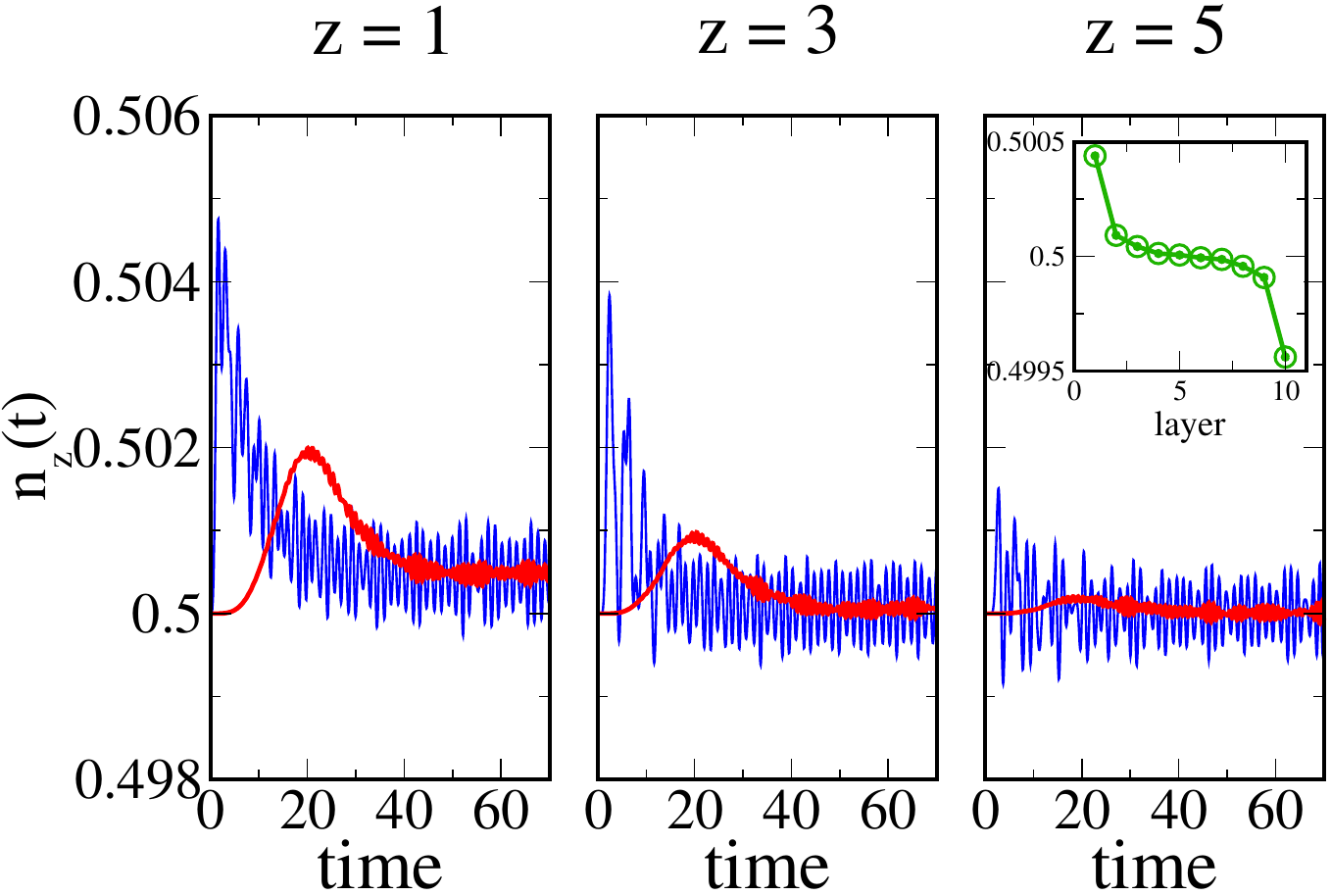}
  \end{center}
  \caption{(Color online) 
    Top: Real-time dynamics of the currents computed at the
    slab-lead contact (thick lines) and between two neighboring layers (light grey lines)
    after a bias quench with  $\Delta V =0.5$ and  $U = 12$.
    for a $N = 10$ slab and different values of the slab-leads coupling $\vhyb$. 
    Inset: Current dynamics for a ramp-like switching protocol 
    $r(t) = [1-3/2\cos(\pi t/\tau_*) + 1/2\cos(\pi t/\tau_*)^3]/4$
    compared 
    to the sudden quench limit ($\tau_*=30$).
    Bottom: Dynamics of the local electronic densities $n_z(t)$ for the
    $1^{\text{st}}$, $3^{\text{rd}}$ and $5^{\text{th}}$ layer and $\vhyb=1.0$.
    Inset: stationary density profile showing an almost flat 
    density distribution with slighlty doped regions near the 
    left and right contacts.
  }
  \label{fig:j_layer}
\end{figure}

We shall now focus on the the dynamics in the presence of an applied bias.
In the Fig.~\ref{fig:j_layer} we report our results for the real-time dynamics
of the currents at the contacts and layers, defined by
Eqs.~(\ref{eq:j_layer})--(\ref{eq:j_contact}), 
after a sudden switch of the bias $\Delta V$ and a flat inner potential.
We observe that the contact and the layer currents display very similar 
dynamics, characterized by a monotonic increase at early times and a
saturation to stationary values at longer times. 
The stationary dynamics displays small undamped oscillations around 
the mean value due to oscillations of the  layer-dependent electronic densities 
(see Fig.~\ref{fig:j_layer}(b)).
As we already mentioned, the persistence of oscillations, \ie the
absence of a true relaxation to a steady-state, is a characteristic of
the essentially mean-field nature of the method. However, this problem can be
overcome either by time-averaging the signal or, as shown in the inset
of Fig.~\ref{fig:j_layer}, using a  finite-time switching protocol $r(t)$
for the voltage bias.
In both cases we end up with the same currents and density profiles, which
result almost flat as a function of the layer, as expected in the metallic case.

\begin{figure}
  \begin{center}
    \includegraphics[scale=0.6]{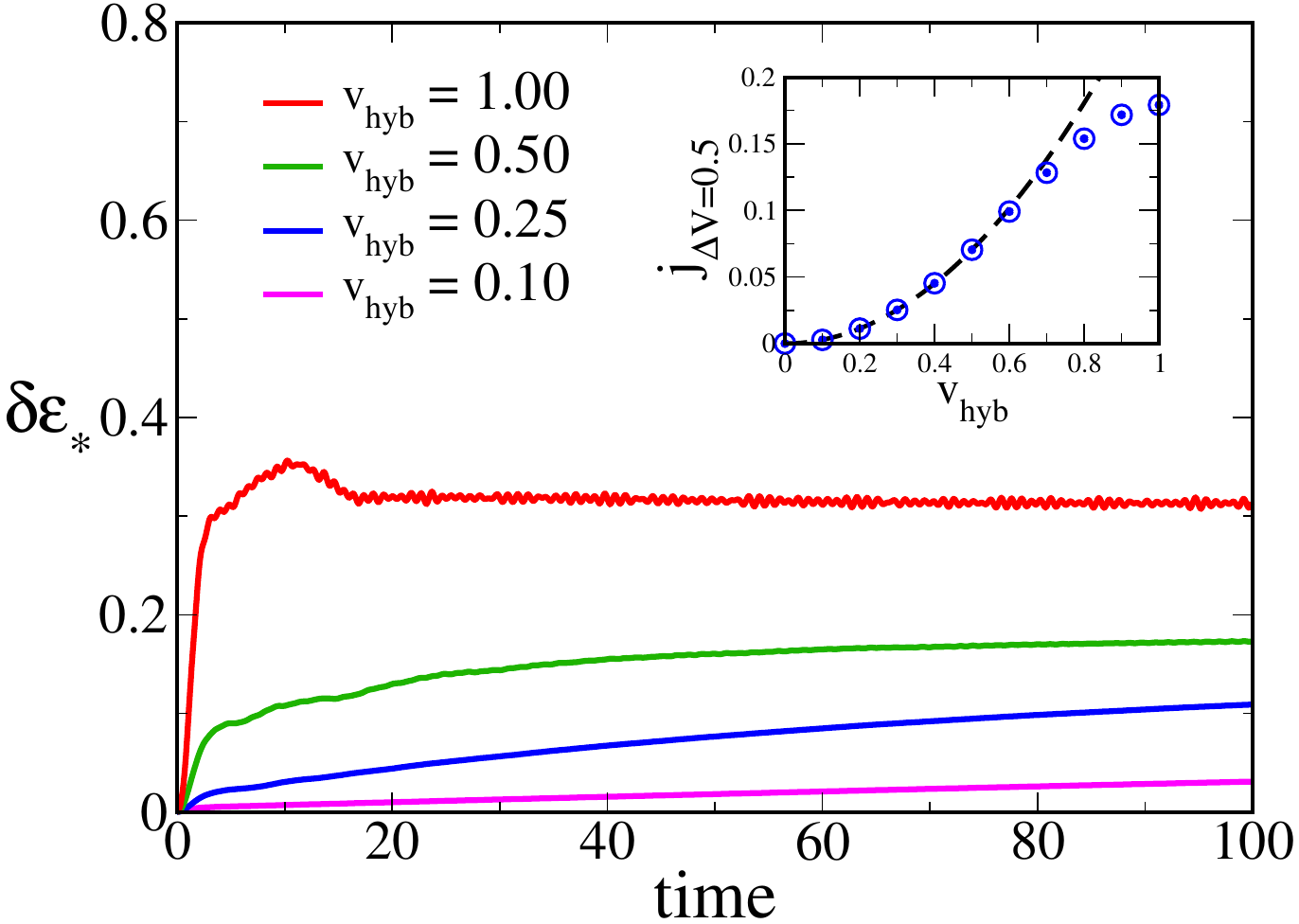}
  \end{center}
  \caption{
    (Color online) 
    Relative variation of the slab internal energy as defined in 
    Eq. \ref{eq:deltae} for the same set of parameters of Fig.
    \ref{fig:j_layer}.
    Inset: Stationary current for $\Delta V = 0.5$ as a function
    of the hybridization with the leads with a fitting curve
    $j(\vhyb) = j_0\, \vhyb^2$ (dashed line).
  }
  \label{fig:ene_slab_dv0.5}
\end{figure}

We highlight that the non-equilibrium dynamics is strongly dependent on the coupling 
between the system (correlated slab) and the external environment (leads),  
represented in this case by the slab-lead tunneling amplitude 
$\vhyb$.
This is evident from the stationary value of the current
that increases as a function of $\vhyb$, as expected since this latter 
sets the rate of electrons/holes injection from the leads into the slab.
Furthermore, the coupling to an external environment is essential to redistribute 
the energy injected into the system after a sudden perturbation so to
lead to a final steady state characterized by a stationary value of
the internal energy.
In order to study the competition between energy dissipation and
energy injection rate we plot, in Fig.~\ref{fig:ene_slab_dv0.5}, the
time-dependence of the relative variation of the slab internal energy with
respect to its equilibrium value:
\begin{equation} 
  \delta \epsilon_*(t) \equiv \frac{E_*(t) - E_*(t=0)}{\left|E_*(t=0)\right|},
 \label{eq:deltae}
\end{equation}
where 
\begin{equation*}
  \begin{split}
    E_*(t) \equiv & \braket{\Psi(t)}{H_{\text{Slab}}}{\Psi(t)} \approx 
    \braket{\Psi_0(t)}{\Hcal_*\big[\hat{\Phi}(t)\big]}{\Psi_0(t)}\\
&    + \sum_z \Tr\left( \hat{\Phi}_z(t)^{\dag} \Hcal_{loc,z} \hat{\Phi}_z(t) 
    \right).
  \end{split}
\end{equation*}
The last expression holds within the TDG approximation.
We observe the existence of two regimes as a function
of the coupling to the leads $\vhyb$. 
When the system is weakly coupled to the external environment the
energy shows an almost linear increase in time without ever reaching
any stationary value. This signals that the dissipation mechanism is
not effective on the scale of the  simulation time. 
For larger values of $\vhyb$, the dissipation mechanism becomes more
effective. The internal energy shows a faster
growth at initial times, due to the larger value of the current setting
up through the system. Further increasing  (see the case $\vhyb\!=\!1.0$ in the
figure) the initial fast rise is of the energhy is followed by a downturn towards
a stationary value, which in turn is reached very rapidly.
As shown in the inset of Fig.~\ref{fig:ene_slab_dv0.5}
the crossover between the non-dissipative and dissipative regimes coincides 
with the point in which the current deviates from linear-response
theory -- which predicts a quadratically  increasing
current $j\!\propto\!\vhyb^2$-- and bends towards
smaller value.

\subsection{Large-bias regime}
\begin{figure}
  \begin{center}
    \includegraphics[scale=0.6]{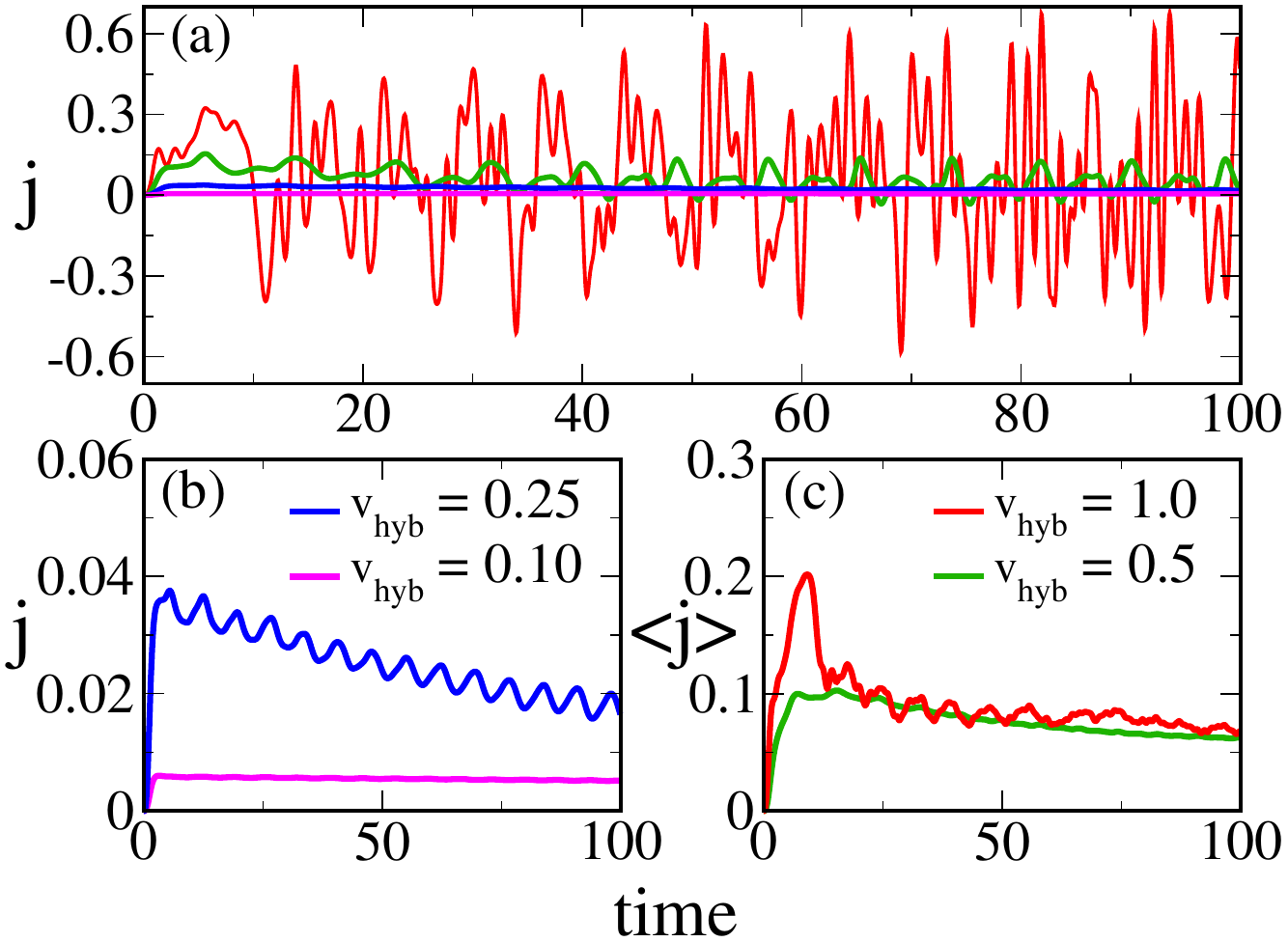}
  \end{center}
  \caption{(Color online) (a) 
    Real-time dynamics for the contact currents
    for the same parameters and values of hybridization coupling
    of Fig.~\ref{fig:j_layer} and $\Delta V=2.0$.
    (b) Blow-up of the currents dynamics for $\vhyb=0.1$ and $\vhyb=0.25$.
    (c) Dynamics of the current time average $\quave{j(t)}$ as defined
    in the main text for $\vhyb=0.5$ and $\vhyb=1.0$.
  }
  \label{fig:j_dv2.0}
\end{figure}

The interplay between the energy injection and the dissipation
highlighted in the dynamics of the slab internal energy (Fig.~\ref{fig:ene_slab_dv0.5})
is a direct consequence of the fact that in our model these two
mechanisms are controlled by the coupling with the same external
environment.
Therefore, we may envisage a situation in which the internal energy of
the slab grows so fast that the leads are unable to dissipate the
injected energy preventing a stationary current to set in. 
This phenomenon occurs at large values of the voltage bias 
($\Delta V\!\gtrsim\!1$) and of the tunneling amplitude 
$\vhyb$, \ie when the slab is rapidly kicked away from equilibrium.  
In order to illustrate this point we report in Fig.~\ref{fig:j_dv2.0}
the current dynamics for the same parameters as in the previous 
Fig.~\ref{fig:j_layer} but for larger value of the voltage bias $\Delta V\!=\!2.0$.
We observe that, while for weak tunneling ($\vhyb\!=\!0.1$) the 
current flows to a steady state, upon increasing $\vhyb$ the
stationary state can not be reached  and strong chaotic oscillations
characterize the long-time evolution. 

\begin{figure}
  \begin{center}
    \includegraphics[scale=0.6]{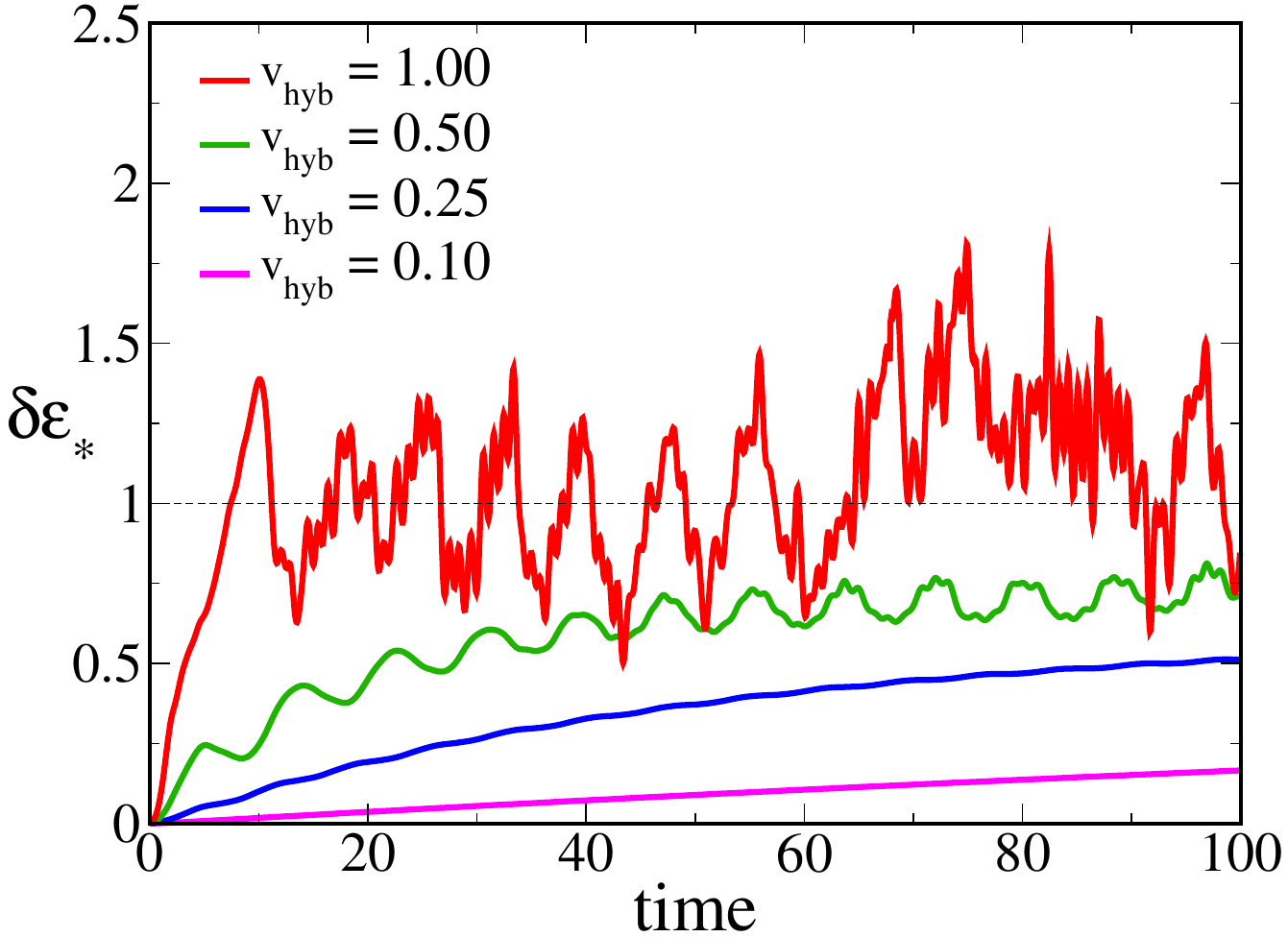}
  \end{center}
  \caption{
    (Color online) Dynamics of the relative energy variation
    for the same parameters in Fig.~\ref{fig:j_dv2.0}.
  }
  \label{fig:ene_slab_dv2.0}
\end{figure}

\begin{figure}
  \begin{center}
    \includegraphics[scale=0.6]{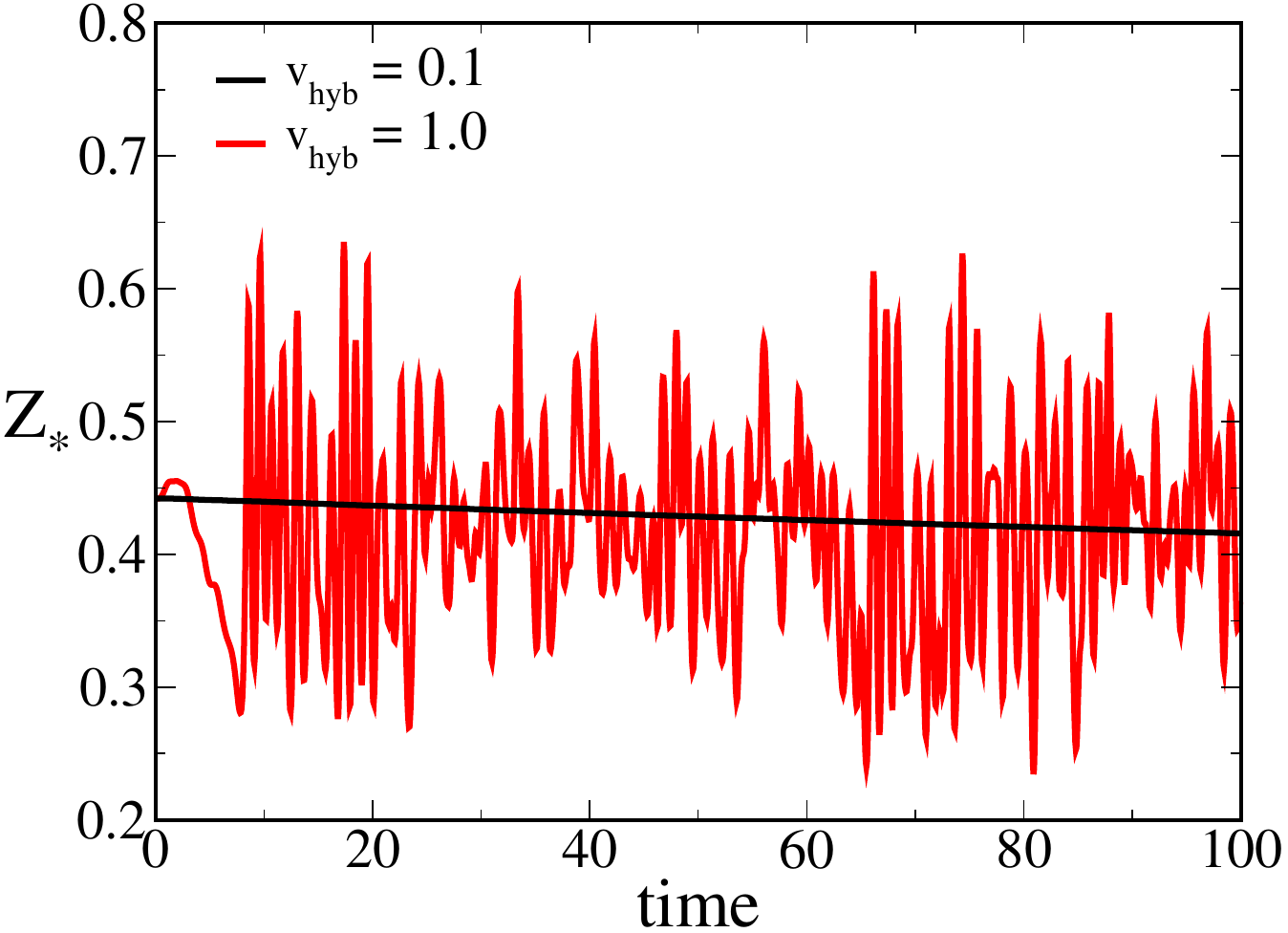}
  \end{center}
  \caption{
    (Color online) 
    Dynamics of the mean quasiparticle weight for the same parameters
    in Fig.~\ref{fig:j_dv2.0} and $\vhyb=0.1$ (black line)
    and $\vhyb=1.0$ (red line).
  }
  \label{fig:Z_dv2.0}  
\end{figure}

Indeed, the inability of reaching a steady-state is intertwined with  
the fast increase of the slab internal energy, as revealed by our
results in Fig.~\ref{fig:ene_slab_dv2.0}.
In particular, for $\vhyb\!=\!1.0$ the relative variation 
of the internal energy rapidly reaches $\delta \epsilon_*(t) \!\approx\! 1$,  
after which it starts to oscillate chaotically just like the currents does.
The same behavior shows up in the dynamics of the quasiparticle weight
averaged over all layers:
\begin{equation}
  Z_*(t) \equiv \frac{1}{N}\sum_{z=1}^N Z_z(t) = \frac{1}{N}\sum_{z=1}^N 
  \left| R_z(t) \right|^2,
\end{equation}
which  displays fast and large oscillations whereas it is
smooth in the case of small $\vhyb$ (see Fig.~\ref{fig:Z_dv2.0}). 

\begin{figure}
  \begin{center}
    \includegraphics[scale=0.6]{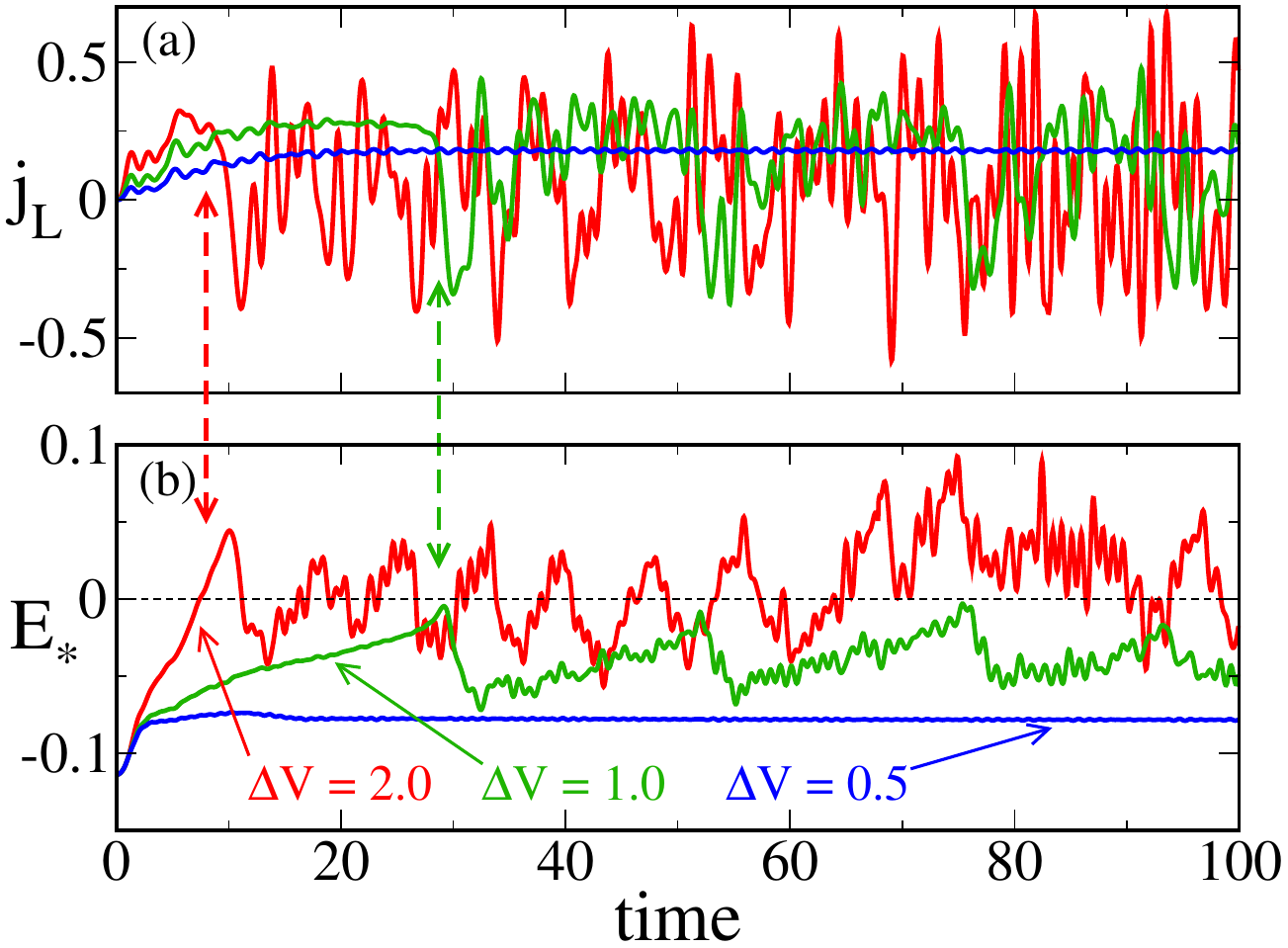}
  \end{center}
  \caption{
    (Color online) 
    Current (a) and internal energy dynamics (b) for $U=12$, $\vhyb\!=\!1.0$
    and three values of the applied bias.
    The occourence of the breakdown of the stationary dynamics
    due to the dynamical transition is highlighted by the vertical arrows.
  }
  \label{fig:j_ene_vk1.0}
\end{figure}

This behavior is similar to that observed across the dynamical
phase-transition in the half-filled 
Hubbard model after an interaction
quench~\cite{SchiroFabrizio_short,WernerPRL} occurring when the
injected energy exceeds a threshold.\cite{Sandri,Giacomo-PRB} 
This correspondence is further supported by noting that the onset of
chaotic behavior occurs precisely when the internal energy $E_*(t)$ of the slab 
reaches zero (see Fig.~\ref{fig:j_ene_vk1.0}). 
The value $E_*\!=\!0$ is indeed the energy of a Mott insulating wavefunction
within the Gutzwiller approximation. This anomalous behavior thus
suggests that as soon as the energy crosses zero $E_*(t)\!\geq\!0$ the
system gets trapped into an insulating state characterized by a
strongly suppressed tunneling into the metal. This prevents
the excess energy to flow back into the leads and does not allow for the relaxation to
a metal with a steady current.

We associate this behavior to a shortcoming of the TDG
approximation, does not include all the
dissipative processes and therefore artificially enhances the
stability of such a metastable state.
If we want to compare this behavior with a real  system, we can argue that the TDG description only
describes a transient state produced by the large initial heating of
the slab that is temporarily pushed into a high-temperature
incoherent phase of the Hubbard model, which takes a long 
time to equilibrate back with the metal leads but evidently not
the infinite time that the TDG approximation suggests. This behavior is
similar to what has been observed by DMFT in the case of an
homogeneous system driven by a static electric field in the absence of
external dissipative channels.~\cite{amaricci_stationary}.

In the case of an interaction quench it was found that, even though
the absence of a true exponential relaxation is faulty, the
time-averaged values of observables as obtained within the TDG
approximation  might still be representative of the true
dynamics.\cite{SchiroFabrizio_short,Sandri} 
This allows us to define a sensible current by time averaging the real-time evolution, \ie through  
\begin{equation}
  \quave{j(t)} = \frac{1}{t}\int_0^t d\tau j(\tau),
  \label{eq:j_ave}
\end{equation}
which indeed approaches a finite value at long enough times (see Fig.~\ref{fig:j_dv2.0}(c)).

\subsection{Current-bias characteristics} 
The overall picture emerging from our investigation of the metallic
case can be summarized by an inspection of the
evolution of the current as a function of the  bias (current-bias
characterisctic) for different values of the interaction strength. 

\begin{figure}[tbh]
\includegraphics[scale=0.6]{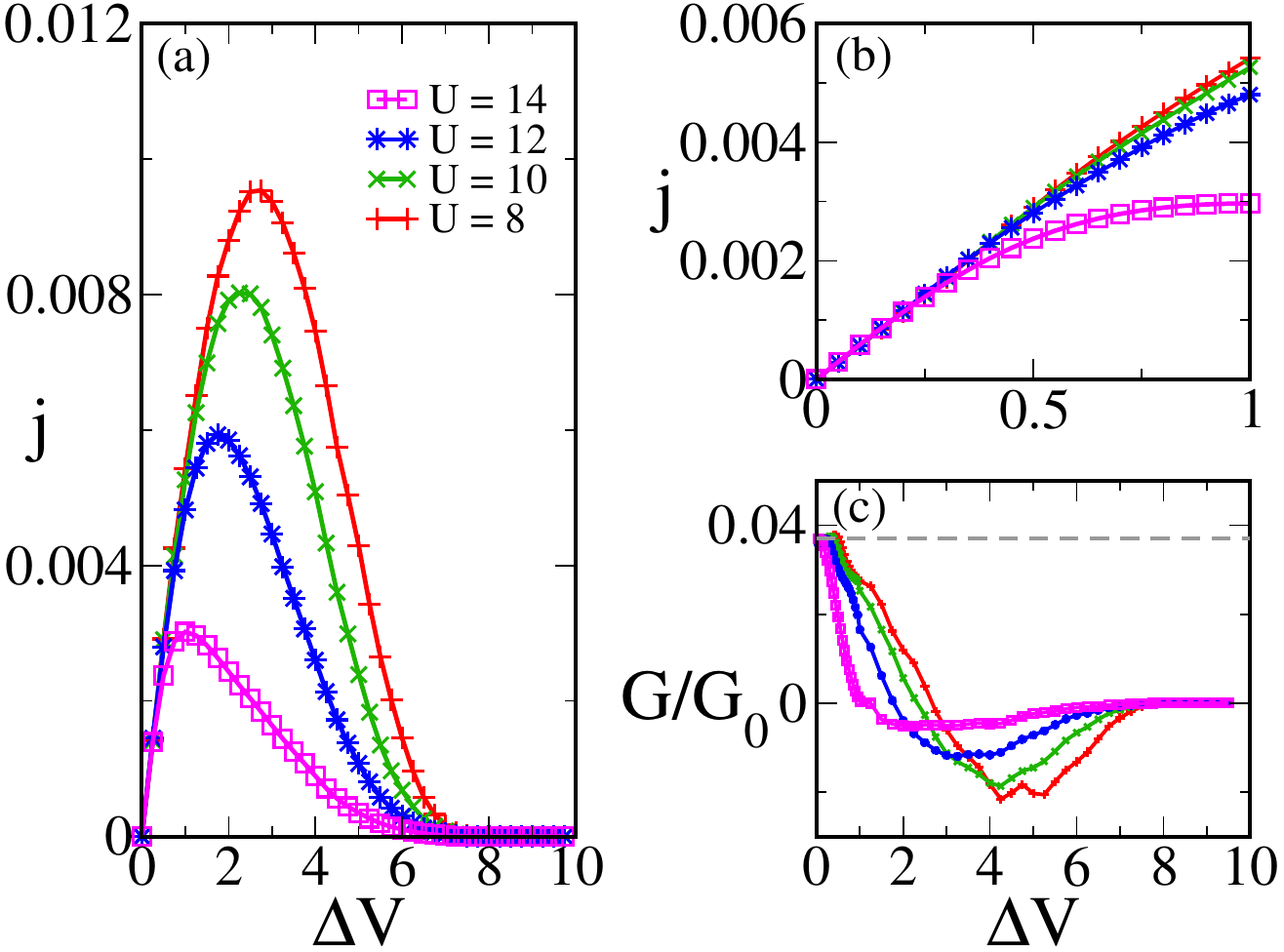}
\caption{(Color online) (a): current-bias characteristics for a $N=10$ 
  slab for different values of $U$ and $\vhyb=0.1$. (b) blow-up 
  of the linear part of the current-bias characteristics.
  (c): Differential conductance measured respect to the quantum conductance
  $G_0 = 1/2\pi$ in our units. Grey line represents the universal
  zero-bias value.
}
\label{fig:iv_universal}
\end{figure}

In the limit of weak coupling to the external environment, we have
seen that the currents display a stationary dynamics in a wide range
of bias values. In Fig.~\ref{fig:iv_universal} we report these
stationary values as a function of the bias for 
$\vhyb\!=\!0.1$ and a wide range of interaction strenghts.
All the curves show a crossover between a linear regime at small bias
and a monotonic decrease for larger values.
This behavior is similar to what was already observed in different contexts.~\cite{
amaricci_stationary,aron_kotliar_dim_crossover,prelovsek_tj_Efield,knap_arrigoni_neqDMFT}
We connect the drop of the current for large biases to the reduction of energy  
overlap between the leads and the slab electronic states at large bias. 
In the linear regime we find that the zero-bias differential conductance 
$G(0)\! =\! \left.\partial j / \partial \Delta V \right|_{\Delta V =0}$ 
is universal with respect to the interaction strength~\cite{lanata_dot_steady,gabi_arxiv}
as expected when the electronic transport is determined only by the low-energy quasiparticle excitations. 

Within the TDG approximation this fact can be easily rationalized by noting 
that quasiparticles are controlled by  the non-interacting Hamiltonian
$\mathcal{H}_*$ in Eq.~(\ref{eq:Hstar}), characterized by a hopping
amplitude renormalized by the factors $|R|\!\leq\!1$.
This leads to an enhancement of the quasiparticle density of states 
by a factor $\nu \!\sim\! 1/|R|^2$ that at  low bias compensates the
reduction of tunneling rate into the leads. 
Conversely, as the bias increases the current-bias characteristics
starts deviating from the universal low-bias behavior and becomes
strongly dependent on the interaction strenght $U$.~\cite{gabi_arxiv} 
In particular, the crossover between the positive and the negative
differential conductance regimes gets shifted towards smaller values
of the bias as $U$ is increased as effect of the shrinking of the
coherent quasiparticle density of states.
\begin{figure}[tbh]
\includegraphics[scale=0.6]{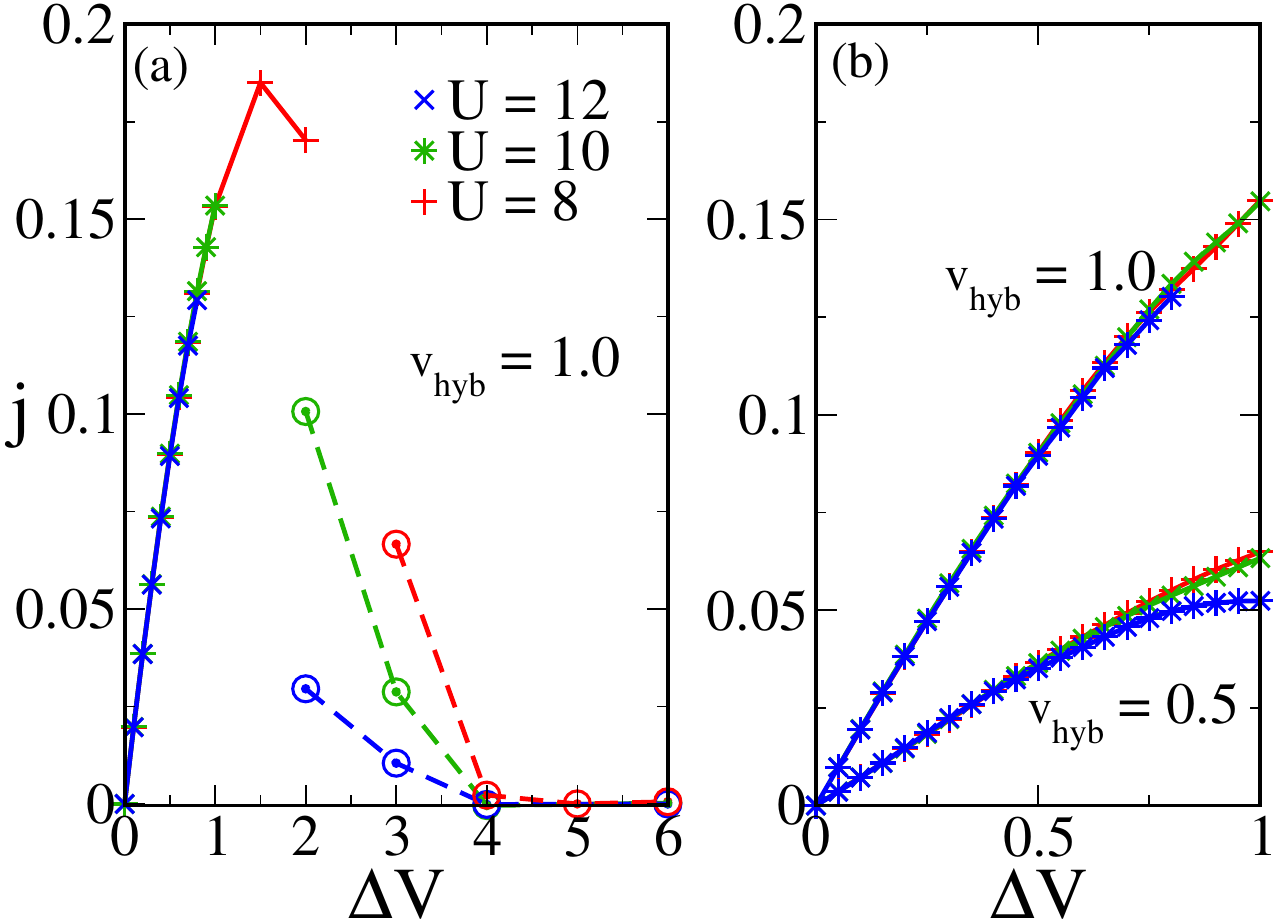}
\caption{(Color online) Left: current-bias characteristics for a $N=10$ 
  slab, different values of $U$ and $\vhyb=1.0$. 
  Plus, cross and star symbols represents stationary currents values, while
  circles represent converged currents time averages values.
  Right: blow-up 
  of the linear part of the current-bias characteristics for $\vhyb=0.5$
  and $\vhyb=1.0$, showing universal zero-bias conductivity
  $G/G_0 \approx 0.452$ and $1.203$ respectively.
}
\label{fig:iv_vk1.0}
\end{figure}

As discussed in the previous section, increasing the coupling 
to the external environment leads to a chaotic regime at large bias,
in a regime where we cannot identify anymore a stationary current. 
However, as mentioned above, we  can still extract 
a meaningful estimate of the current through its time-average Eq.~\eqn{eq:j_ave}), 
restricting to the range of bias for which the latter is well converged.
This is explicitely illustrated in Fig.~\ref{fig:iv_vk1.0} for the current-bias
characteristics at $\vhyb\!=\!1.0$. The open circles
represent currents computed using converged time averaged 
while the other symbols represent currents characterized by a stationary dynamics.
Our results show that the curves have qualitatively the same features of the small 
$\vhyb$ case with a universal linear conductance and a crossover to a
negative conductance regime.

\begin{figure}
  \begin{center}
    \includegraphics[scale=0.6]{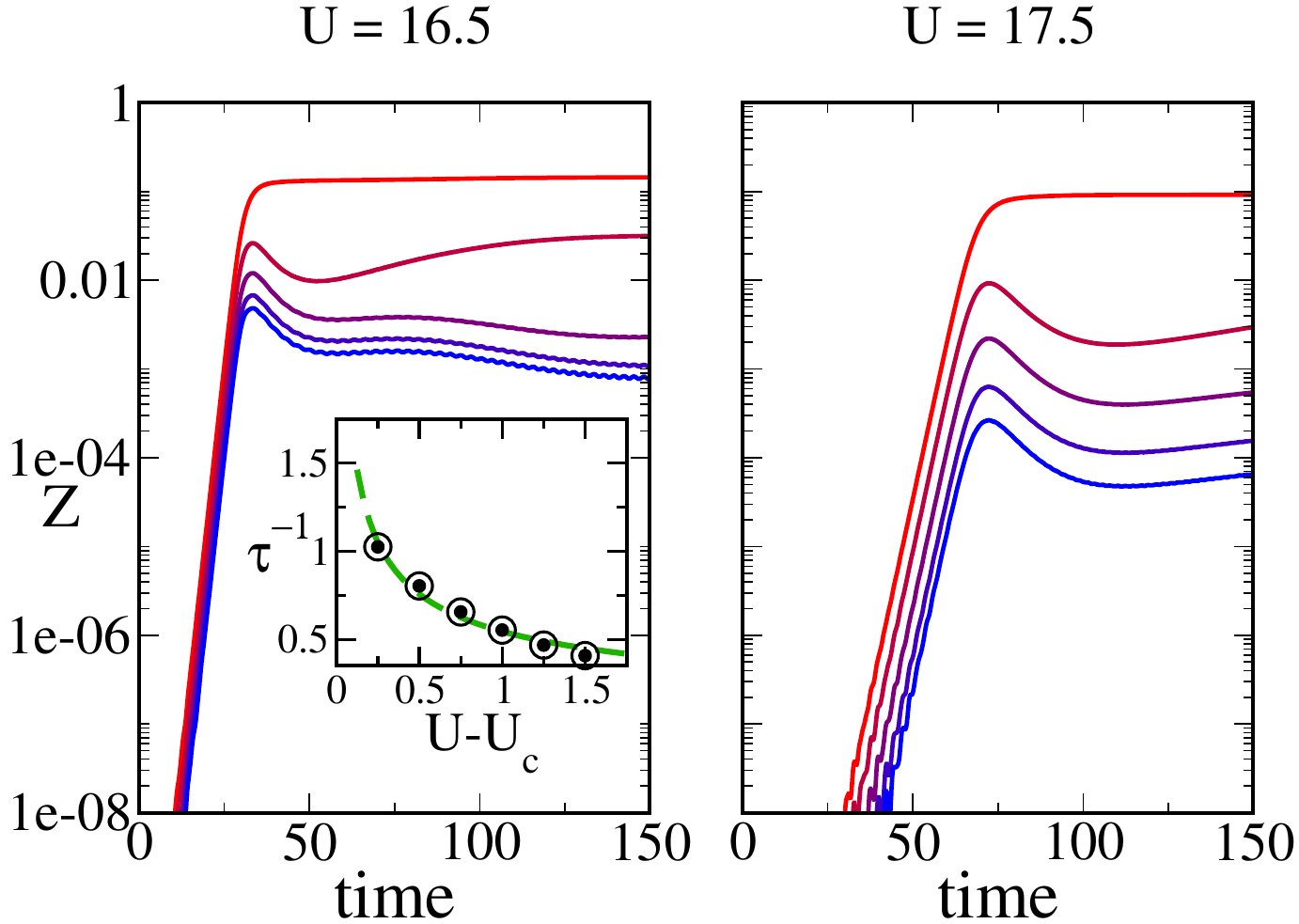}
  \end{center}
  \caption{(Color online) Real-time dynamics for the quasiparticle
    weights from layer $1$ to $5$ (from top to bottom) of a $N=10$ 
    Mott insulating  slab suddenly coupled to the metallic leads ($v_{\text{hyb}}=1.0$).
    $U=16.5$ and $U=17.5$. 
    Inset: inverse of the characteristic time for the exponential 
    quasiparticle formation $\tau^{-1} \sim (U-U_c)^{-\nu_*}$, $\nu_* \approx 0.4753$.
  }
  \label{fig:wake_up_ins}
\end{figure}

\section{Dielectric breakdown of the Mott insulating phase}
\label{sec:noneq_insulator}
We now move the discussion to the effect of an applied voltage bias
to a slab which is in a Mott insulating regime because $U > U_c$. 
Unlike the metallic case, we now assume that the field  
penetrates inside the slab, leading to a linear potential profile of
the form $E_z=\Delta V  (N+1-2z)/2(N+1)$ matching the chemical
potential of the  left and right
leads for $z=0$ and $z=N+1$ respectively.

\subsection{Evanescent bulk quasiparticle}
\begin{figure}
  \begin{center}
    \includegraphics[scale=0.6]{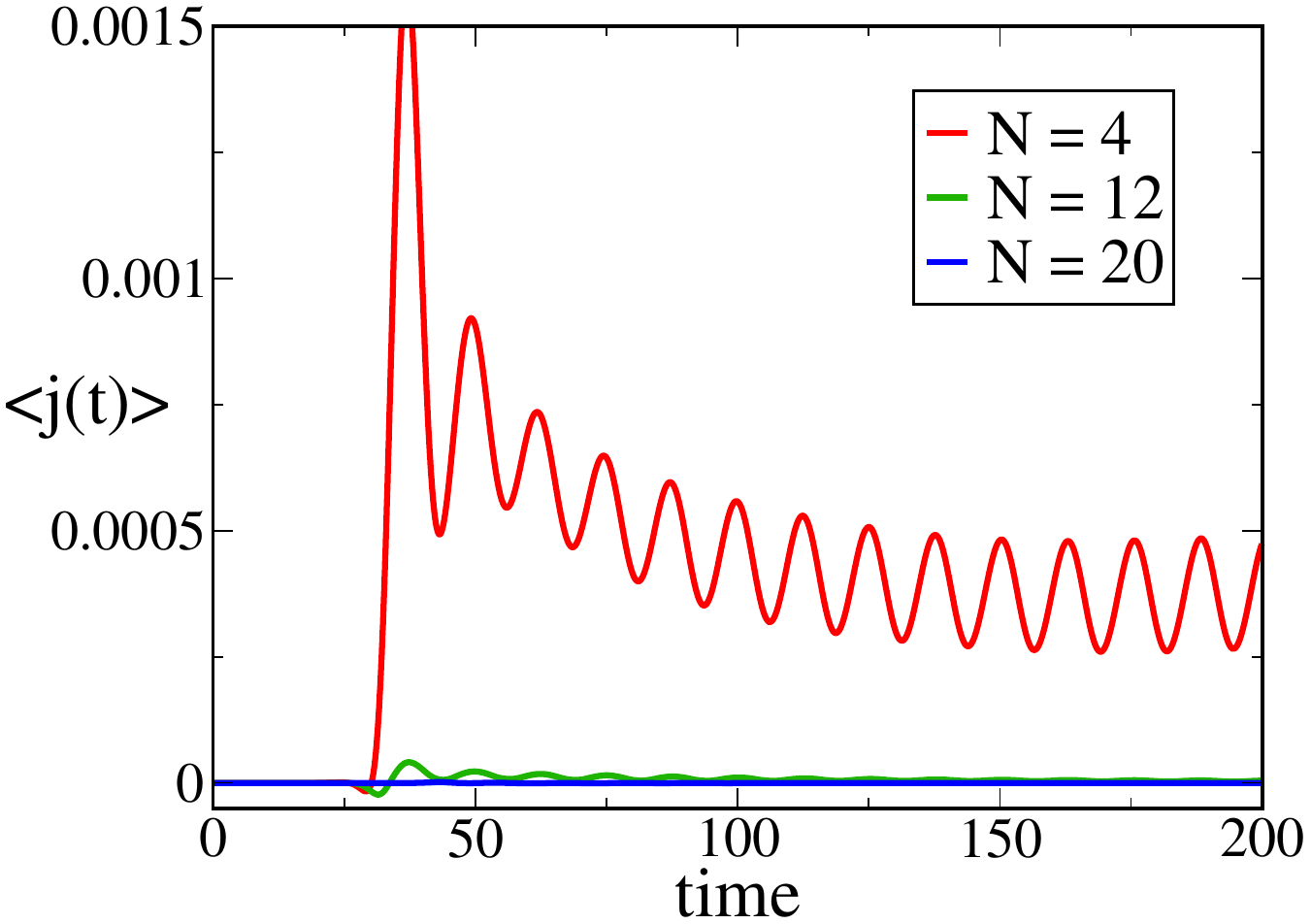}
    \caption{(Color online) 
      Time-averaged currents for three slabs with applied bias $\Delta V =0.5$.
      $U=16.5 > U_c$, $\vhyb = 1.0$ and $N=4,12,20$ (from top to bottom).}
    \label{fig:j_mottL4_12}
  \end{center}
\end{figure}

Within the Gutzwiller approximation the Mott insulator is 
characterized by a vanishing number of doubly 
occupied and empty sites as well as by a zero renormalization factor
$R\!=\!0$, leading to a trivial state with zero energy.
However, it has been shown that in the presence of the metallic leads {\it evanescent} quasiparticles
~\cite{Borghi_prb10,freericks_fragileFL_slab}
appear inside the insulating slab. This is revealed by a finite
quasiparticle weight which is maximum at the leads and decays exponentially in the bulk of the slab with a
characteristic length $\xi\sim (U-U_c)^{-1/2}$ which defines
the critical correlation length of the Mott transition.~\cite{Borghi_prb10} 

In Fig.~\ref{fig:wake_up_ins} we show the dynamics of the formation of {\it
  evanescent} quasiparticles after the sudden switch on of the
coupling to the leads $\vhyb$. 
We observe a rapid increase of the
quasiparticle weight as soon as the coupling is switched on. The rapid
increase can be reasonably well parameterized as an exponential with a
characteristic growth time $\tau$. The results for $\tau^{-1}$
reported in the inset of the left panel of Fig.~\ref{fig:wake_up_ins})
clearly show that the increase of the quasiparticle weight becomes
faster as the Mott transition is approached.
Interestingly, the exponential growth is not limited to the boundary
layers close to the leads, but it is present throughout the slab, with
a characteristic time $\tau(z)$ which is nearly uniform in space.

Such an exponential growth is suggestive of an avalanche effect,
driven by the combined action of the high-energy excitations (Hubbard
bands) and of the quasiparticles,
which within the Gutzwiller approach can be associated to the
variational parameters $\Phi_{z,n}(t)$ and to the non-interacting
Slater determinant $\mid\!\Psi_0(t)\rangle$, respectively. 

As outlined in Appendix \ref{Growth of the living layer}, we can
reproduce the long-time approach to the steady state corresponding 
to evanescent quasiparticles at equilibrium, considering a simplified dynamics
in which we neglect the dynamics of the Slater determinant $\mid\!\Psi_0(t)\rangle$
and take into account only that of $\Phi_{z,n}(t)$. The latter can be analytically written
in terms of a Klein-Gordon-like equation for the hopping renormalization factors $R(z,t)$
\be
\fract{1}{c^2}\,\ddot{R}(z,t) - \nabla^2\,R(z,t) + m^2\,c^2\,R(z,t) = 0,
\label{KG}
\ee
with parameters (see Appendix \ref{Growth of the living layer}):
\be
c^2 = \fract{U}{24},\qquad m^2 c^2 = 6(U-1) = \xi^{-2}.
\label{parameterKG}
\ee
As anticipated above the simplified dynamics described by Eq.~\ref{KG}  correctly captures the
long-time behaviour of the system, but it can not reproduce the short
time exponential growth. 
In the latter regime the time evolution is indeed  governed by the
interplay between Hubbard bands and quasiparticles, responsible for the evenescent quasiparticle formation
into the Mott insulating slab, which is neglected in the approximation
leading to Eq.~\ref{KG}.

The presence of the evanescent bulk quasiparticle provides a conducting channel
accross the slab, possibly leading to finite currents upon the application of a
finite bias.
In particular, we expect that if the slab lenght smaller than the
decay length $\xi$ every finite bias $\Delta V$ is sufficient to induce
a finite current through  the slab. On the other hand we expect the
current to be suppressed when the slab is longer than $\xi$
This is confirmed by the results reported in Fig.~\ref{fig:j_mottL4_12} 
where we show the average current for a bias $\Delta V=0.5$, 
in the linear regime in the metallic case, and different slab sizes $N$. 
A finite current is rapidly injected for
small $N\!=\!4$, whereas it does not for larger systems 
({\it e.g.} $N\!=\!12$ or $N\!=\!20$).  

\subsection{Dielectric breakdown currents}
\begin{figure}[t]
  \includegraphics[scale=0.6]{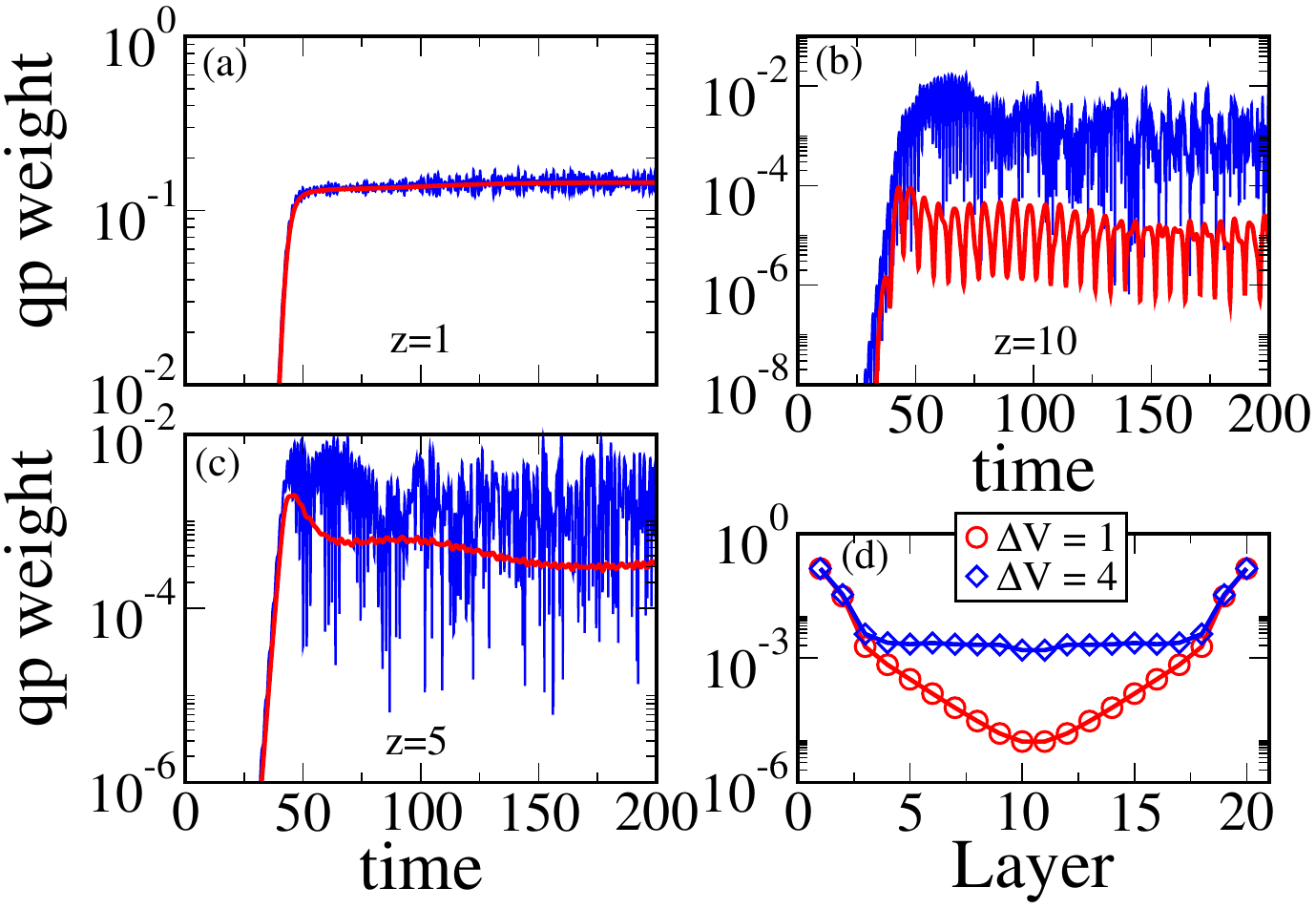}
  \caption{(Color online) Layer dependent quasiparticle weight dynamics
    for layers 1 (a), 5(b) and 10(c) of a $N=20$ slab, $U = 16.5$, $\vhyb=1.0$
    and two values of the applied bias $\Delta V=1.0$(red lines) and $\Delta V=4.0$
    (blue lines). (d): time-averaged stationary quasiparticle weight profile.
  }
  \label{fig:z_mott}
\end{figure}
Increasing the value of the applied bias we observe an enhancement of the 
quasiparticle weight throughout the slab. This effect is illustrated
in Fig.~\ref{fig:z_mott} where panels (a-c) show the the dynamics of the 
quasiparticle weights in a driven Mott insulating slab with different 
values of the bias for three different layers ($z=1, 5, 10$).
While the dynamics is characterized by strong oscillations 
reminescent of the inchoerent dynamics discussed in Sec.~\ref{sec:noneq_metal}
for the metallic slab under a large applied bias, 
the time-averaged quantities in the long-time limit converge to
stationary values. The spatial distribution as a function of the layer
index shows a strong enhancement in the bulk upon increasing the bias (Fig.~\ref{fig:z_mott}d).

\begin{figure}
  \includegraphics[scale=0.6]{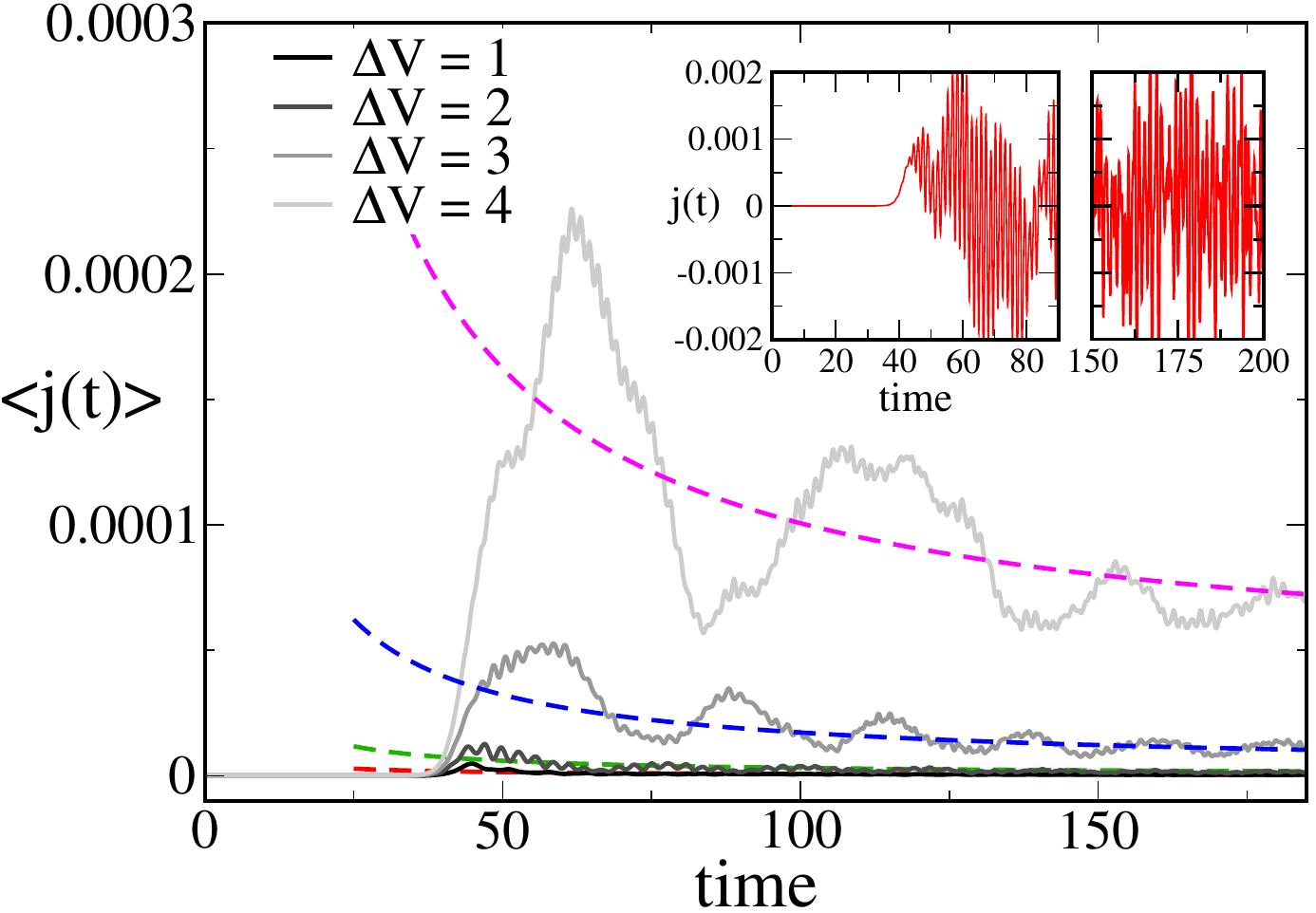}
  \caption{(Color online) 
    Time-averaged currents for the same parameters in Fig.~\ref{fig:z_mott}
    and $\Delta V=4.0,3.0,2.0$ and $1.0$ (from top to bottom). 
    Dashed lines are fitting curves from Eq.~(\ref{eq:jfit}).
    Inset: Real time dynamics of the current for $\Delta V = 4.0$.}
  \label{fig:j_exe_mott}
\end{figure}

Such enhancement results in a finite current flowing.
Indeed, as shown in Fig.~\ref{fig:j_exe_mott}, the time-averaged
current has a damped oscillatory behavior that  converges towards a
steady value, although  the real-dynamics follows a seemingly chaotic pattern
(see the inset). We extract the stationary values by fitting the current time-averages
with:
\begin{equation}
  \quave{j(t)} = j_{\text{steady}} + \frac{\alpha}{t}.
  \label{eq:jfit}
\end{equation}
As evident by looking at the results reported in Fig.~\ref{fig:j_exe_mott}, the
stationary value of the current has a non-linear behavior as a
function of the applied bias. This effect can be better appreciated in the next
Fig.~\ref{fig:iv_mott}, where we plot the current-voltage characteristics for
increasing values of the slab size $N$.

Interestingly, the current displays an exponential activated behavior
with a characteristic threshold bias which is well described by
\begin{equation}
  j_{\text{steady}}(\Delta V) = \gamma\, \Delta V \,\text{e}^{-\Delta V_\text{th}/\Delta V}.
  \label{eq:iv_mott_fit}
\end{equation}
Fitting the data with the above relation we obtain $\Delta V_\text{th}\propto
N$ (see inset in Fig.~\ref{fig:iv_mott}) so that we can rewrite
Eq.~\eqn{eq:iv_mott_fit} as function of the electric field $\Delta
V/N$:
\begin{equation}
  \fract{j_{\text{steady}}(\Delta V)}{\Delta V} = \gamma\, \text{e}^{-E_\text{th}/E},
  \label{eq:iv_mott_fit-2}
\end{equation}
which introduces a size-independent threshold field $E_\text{th}$. 
This expression is suggestive of a Landau-Zener type of dielectric
breakdown\cite{oka_prl2003,oka_prl2005}, similar to the results obtained within DMFT studies
of either homogeneous~\cite{eckstein_dielectric_breakdown}  
and inhomogeneous systems.~\cite{SOkamoto_PRB,Eckstein_slab}

\begin{figure}
  \includegraphics[scale=0.6]{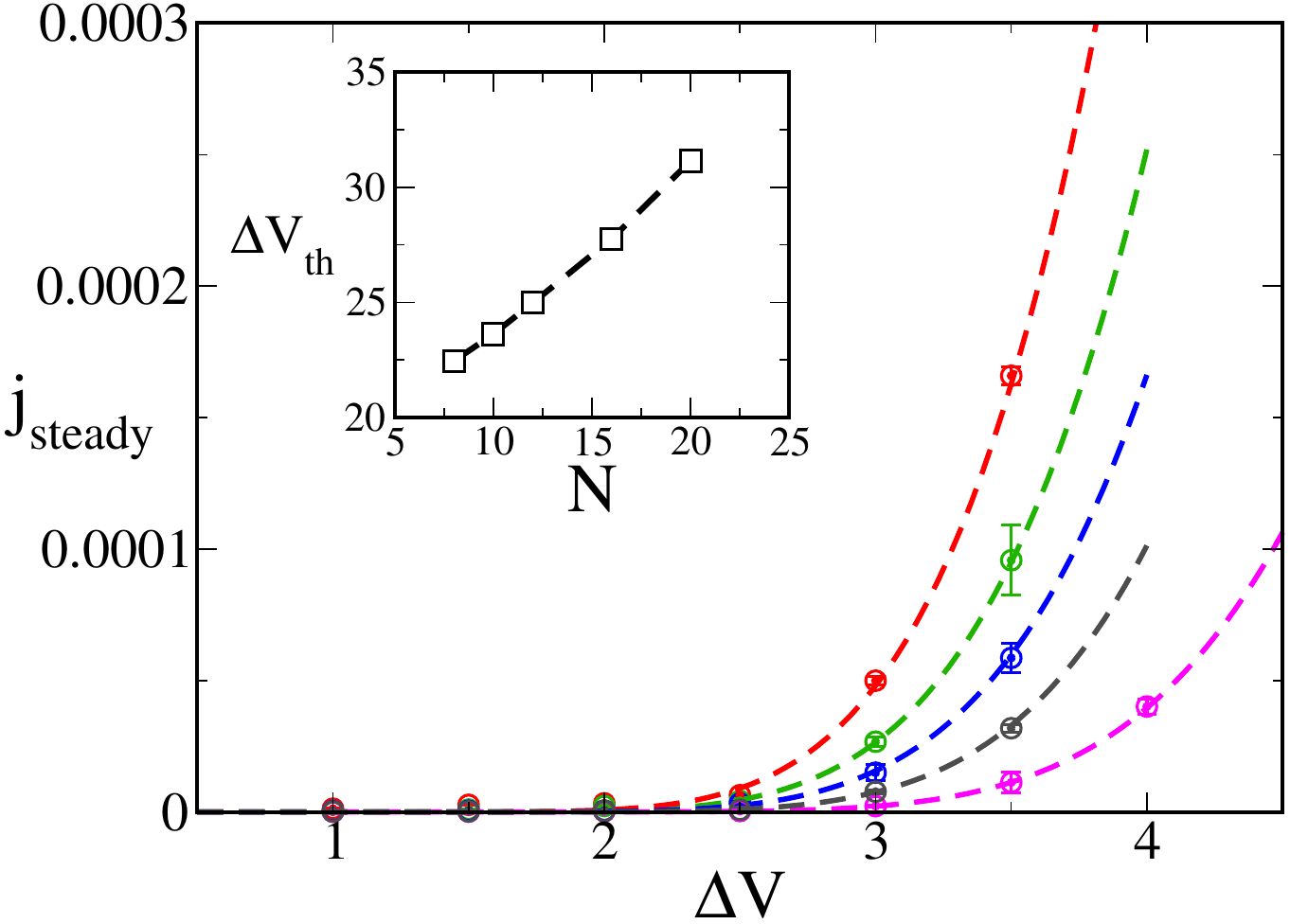}
  \caption{(Color online) Current bias characteristics for $U=16.5$ and 
    different values of the slab lenght $N=8,10,12,16$ and $20$ 
    (from top to bottom).
    Dashed lines represent fitting curves with Eq.~(\ref{eq:iv_mott_fit}).
    Insets: threshold bias $\Delta V_{th}$ as a function of the slab size.
  }
  \label{fig:iv_mott}
\end{figure}

The Gutzwiller scenario for the dielectric breakdown is further
supported by the simple calculation for the stationary regime outlined in the Appendix
\ref{app:Landau-Zener}, which follows the analysis reported in
Ref.~\onlinecite{Borghi_prb10} for the equilibrium case, considering a
single metal-Mott insulator interface in the presence of an
electrochemical potential $\mu(z)$.
As detailed in \ref{app:Landau-Zener}, we find that for weak $\mu(z)$, the 
hopping renormalization factor $R(z)$ satisfies the equation
\be
\nabla^2 R(z) = \Big(m^2 c^2 - \frac{2\mu(z)^2}{c^2}\Big)\,R(z),\label{eq:LZ}
\ee
which is nothing but the stationary Klein-Gordon equation \eqn{KG} in
the presence of a field, or alternatively, the Schr{\oe}dinger
equation of a particle impinging on a potential barrier. 
For a constant electric field $\mu(z)\!=\!E\,z$ and within the
WKB approximation, we obtain the stationary transmission probability beyond
the turning point $z_*$ of the barrier  (see Appendix
\ref{app:Landau-Zener}):
\be
\left| R(z>z_*)\right|^2 \sim \exp\left(-\fract{E_\text{th}}{E}\right),\label{transmission}
\ee
where 
\be
E_\text{th} = \fract{\pi}{\sqrt{8}}\;m^2\,c^3\sim \xi^{-2}.\label{E_th}
\ee
This calculation identifies the transmission probability (Eq.~\ref{transmission}) 
with the dielectric breakdown currents (Eq.~\ref{eq:iv_mott_fit-2}) and predicts via the definition
of the correlaton lenght (Eq.~\ref{parameterKG}) a threshold electric field increasing 
with the interaction strenght.~\cite{eckstein_dielectric_breakdown}

\subsection{Quasiparticle energy distribution}
Inspired by the evidence that in our description the transport activation is driven 
by an enhancement of the bulk quasiparticle weight [see  
Fig.~\ref{fig:z_mott}(d)] in this section we focus on the spatial 
distribution of the quasiparticle energy throughout the slab.
In order to estimate the time evolution of the quasiparticle energy
levels we compute the time-evolution of the layer-dependent chemical
potential in the effective non-interacting model Eq.~\ref{eq:Hstar},
introduced by the coupling to external voltage bias. 
This quantity can be easily extracted by means of the following unitary 
transformation of the uncorrelated wavefunction:
\begin{equation}
  \ket{\varphi_0(t)}\!\equiv\! \mathcal{U}(t)\,\ket{\Psi_0(t)}\,,\;
  \mathcal{U}(t)\! =\! \prod_{\br,z}\exp{\Big[i \lambda_z(t)\, \hat{n}_{\br,z} \Big]}
  \label{eq:unitary-1}
\end{equation}
where $\lambda_z(t)$ is the time-dependent phase of the hopping
renormalization parameters $R_z(t)\! \equiv \!\rho_z(t)\text{e}^{i
  \lambda_z(t)}$, with real $\rho_z(t)\!\geq\!0$. 
Substituting Eq.~\eqn{eq:unitary-1} into  Eq.~(\ref{eq:dot_psi}) we
obtain a transformed Hamiltonian that now contains only real hopping
amplitudes at the cost of introducing a time-dependent local chemical
potential terms $\mu_*(z,t)$, namely:
\begin{equation}
  i\partial_t \ket{\varphi_0(t)} = h_*(t)\; \ket{\varphi_0(t)}
\end{equation}
where the effective Hamiltonian reads:
\begin{equation}
   \begin{split}
 h_*(t) &= H_{\text{Leads}} +  \sum_{z=1}^N \sum_{\bk,\sigma}\, \rho_z(t)^2 \, \epsilon_{\bk}\, \dc_{\bk,z,\sigma} \da_{\bk,z,\sigma} \\
  &+ \sum_{z=1}^{N-1} \sum_{\bk,\sigma}\Big( \rho_{z+1}(t)\,\rho_z(t) \; \dc_{\bk,z+1,\sigma} \da_{\bk,z,\sigma} + H.c.\Big) \\
  &+ \sum_{\alpha=L,R}\, \sum_{\bk, \kp, \sigma}\Big(v_{\kp} \rho_{z_{\alpha}}(t) \cc_{\bk \kp \alpha  \sigma} \da_{\bk z_{\alpha} \sigma }  + H.c.\Big) \\
  &+ \sum_{z=1}^{N} \sum_{\bk,\sigma}\, \mu_*(z,t)\,  \dc_{\bk,z,\sigma} \da_{\bk,z,\sigma},
  \end{split}
  \label{eq:hstar}
\end{equation}
and with $\mu_{*}(z,t)\! =\! \frac{\partial}{\partial t}\lambda_z(t)$ that 
plays the role of an effective chemical potential for the
quasiparticles under the influence of the bias.

\begin{figure}
  \includegraphics[scale=0.6]{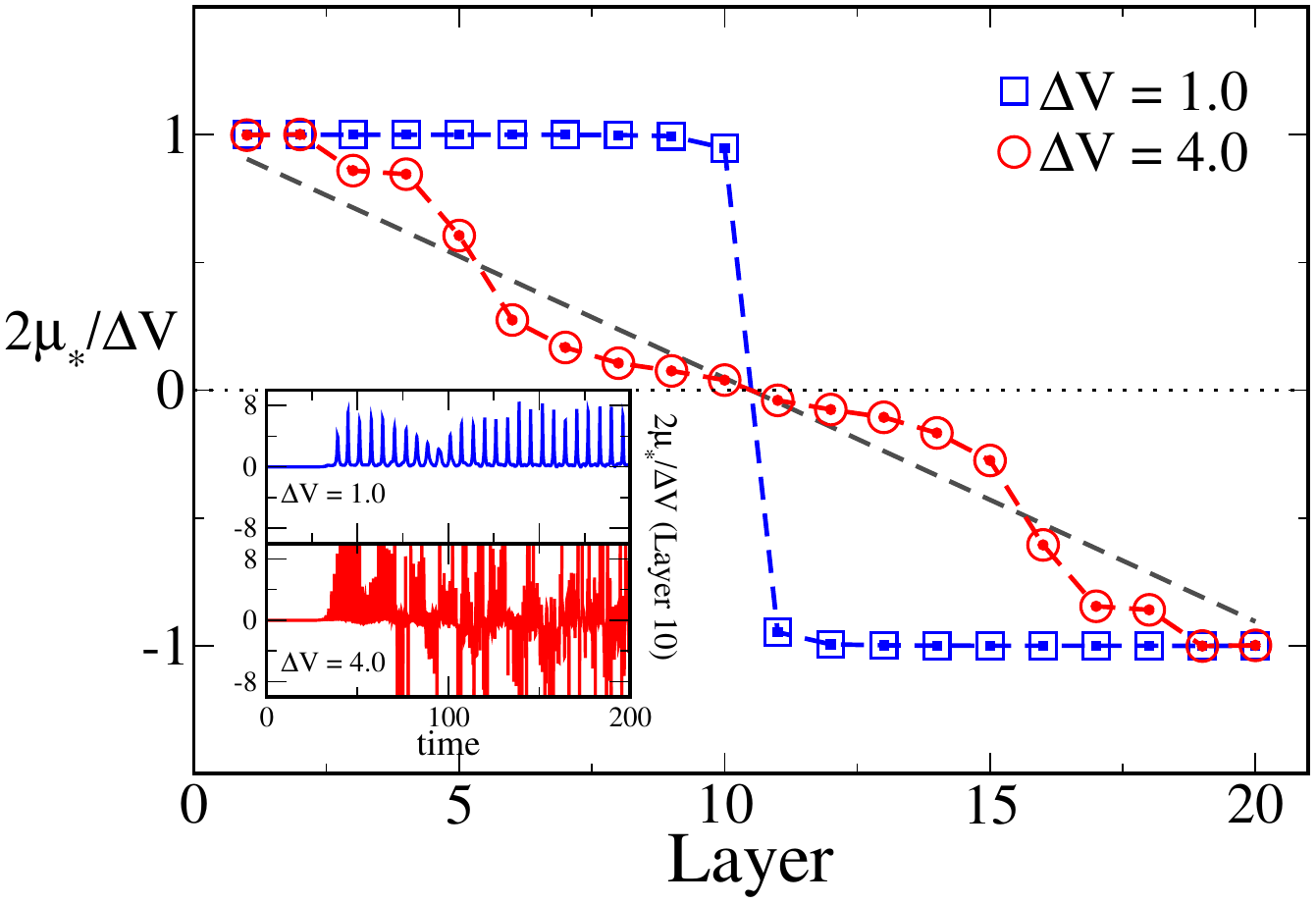}
  \caption{
    (Color online)     
    Layer-dependent quasiparticle effective chemical potential profile.
    Parameters are the same of Fig.~\ref{fig:j_exe_mott} for $\Delta V=1$ 
    and $\Delta V=4$.
    The grey dashed line represents the applied bias linear profile.
    Inset: Real-time dynamics of the quasiparticle chemical potential on the 
    $10^{\text{th}}$ layer.
    All data are plotted with respect to the leads' chemical potential
    absolute value $\Delta V/2$.
  }
  \label{fig:effective_qp_level}
\end{figure}

The time-average of this quantity in the long-time regime we obtain
the energy profile as a function of the position in the slab of the
stationary quasiparticle effective potential, reported in 
Fig.~\ref{fig:effective_qp_level}, locating the energies of the
quasiparticles injected from the leads into the slab. 
As expected, for any value of the applied voltage bias the quasiparticles near the boundaries
are injected at energies equal to the chemical potentials of the two leads, 
\ie $\mu_*\!=\!\pm\Delta V/2$. On the other hand, the behavior inside
the bulk of the slab depends strongly on the value of the applied
bias.

At a small bias, represented in Fig. ~\ref{fig:effective_qp_level}  by $\Delta V\! =\!1$, a value corresponding to an
exponentially suppressed current, the chemical potential remains
essentially flat as the bulk is approached from any of the two leads,  despite the presence of a linear potential drop $E_z$.
This gives rise to a step-like chemical potential profile
with a jump $\Delta \mu_* \!\approx\! \Delta V $ at the center of the slab.
The presence of this jump suppresses the overlap between 
the quasiparticle states on the two sides, preventing the tunneling from the left
metallic lead to the right one and ultimately leading to an
exponential reduction of the current.

On the opposite limit of a large enough bias (\eg $\Delta V\! =\!4$) a
finite current flows through the slab, corresponding to a smoother
profile of effective chemical potentials. Indeed, in the bulk
$\mu_*(z)$ takes a weak linear drop behavior as expected for a metal, 
and slightly reminiscent of the applied linear potential drop $E_z$. 
In this regime the large overlap between quasiparticle states near the
center of the slab allows quasiparticle to easily tunnel from the left to the 
right side, giving rise to a finite current as outlined in the
previous Fig.\ref{fig:j_exe_mott}.

The disappearence of the effective chemical potential discontinuity in
the middle of the slab for large bias is determined by the presence of
strong oscillations of this quantity between positive and negative
values, as shown in the inset of Fig.~\ref{fig:effective_qp_level}.
This suggests that, even though the the quasiparticle chemical potential
averages to an almost zero value at very long times, the
quasiparticles dynamically visit electronic states far away from the local Fermi energies. 
We interpret this behaviour as the signal of a strong feedback 
of the dynamics of the local degrees of freedom
Eq.~(\ref{eq:dot_phi}) onto the the quasiparticle evolution, due to
the proximity of a resonance between quasiparticles and the incoherent
Mott-Hubbard side bands.
Interestingly, even though in our descpription there is no high-energy incoherent
spectral weight, this scenario is reminiscent of the formation of 
coherent quasiparticle structures inside the Hubbard bands as observed in 
previous studies using steady-state formulation of non-equilibrium DMFT.~\cite{SOkamoto_PRB}

\section{Conclusions}
\label{sec:conclusions}
We used the out-of-equilibrium extension 
of the inhomogeneous Gutzwiller approximation to study the dynamics of a correlated slab contacted to metal leads 
in the presence of a voltage bias.
On one side this allowed us to investigate the non-equilibrium counterpart
of known interface effects arising in strongly correlated heterostructures,
such as the {\sl dead} and {\sl living} layer phenomena. On the other
we studied the non-linear electronic transport of quasiparticles 
injected into the correlated slab under the influence of an applied
bias. 
 
In the first part of the paper we considered a slab in a
metallic state in the absence of the bias, when the correlation
strength is smaller than the critical value for a Mott transition.  Initially we focused on  the zero-bias regime and studied the 
spreading of the doubly occupied sistes injected into the slab after a sudden switch
of a tunneling amplitude with the metal leads. 
Specifically we found a ballistic propagation of the perturbation 
inside the slab, leaving the system in a stationary state equal to the equilibrium
one, with an excess of double occupancies concentrated near the contacts
and a consequent enhancement of the quasiparticle weight at the
boundaries of the slab.
We characterized this ``awakening'' dynamics of the {\sl living} layer from
the initial {\sl dead} one in terms of a characteristic time-scale
which  diverges at the Mott transition. This divergence
allow us to  identify this timescale  as the dynamical counterpart of the equilibrium
correlation lenght $\xi$.~\cite{Borghi_prl09}

In the presence of a finite bias we studied the conditions for the formation 
of non-equilibrium states, characterized by a finite current flowing
through the correlated slab.
We demonstrated that this process is strongly dependent on the coupling 
with the external environment represented by the biased metal leads,
which at the same time act as the source of the non-equilibrium 
perturbation and as the only dissipative channel. 
For weak coupling between the leads and the slab we found stationary currents flowing
in a wide range of bias. 
Conversely for large couplings we identified a strong-bias regime 
in which the system is trapped into
a metastable state characterized by an effective  slab-leads
decoupling. 
This is due to an exceedingly fast energy increase and to the lack of 
strong dissipative processes in the Gutzwiller method, which prevents 
the injected energy to flow back into the 
leads and the current to reach a stationary value. 
Studying the current-bias characteristics in the range of parameter for which 
the system is able to reach a non equilibrium stationary state,
we observed a crossover from a low-bias linear regime, which we find 
universal with respect to the interaction $U$, 
to a regime with negative differential conductance typical of finite bandwith systems.
Considering suitable long-time averages of the current  
we have been able to observe the same phenomenology in the region of parameters
for which, due to the aforementioned anomalous heating,
the current dynamics does not lead to an observable stationary value.

In the second part of this work we turned our attention to the
dynamical effect of a bias on a Mott insulating slab, when the
interaction strength exceeds the Mott threshold. 
Following the analysis carried out in the metallic case, we  considered
the formation of a evanescent bulk quasiparticles after a sudden 
switch of the slab-leads tunneling amplitude in a zero-bias setup. 
In this case, we have found that the {\sl living} layer formation is accompanied
by an exponential growth of the quasiparticle weight, suggestive of an 
avalanche effect determined by the interplay between the dynamics 
of the quasiparticles and the local degrees of freedom.

In the presence of a finite bias, we studied the conditions under which
these evanescent quasiparticles can lead to the opening of a conducting channel through the
insulating slab. 
We showed that at very low bias this is the case only for a very small slab, for which 
the correlation length $\xi$ is of the same order of the slab size. 
For larger samples we found that the currents are exponentially 
activated with a threshold bias $\Delta V_\mathrm{th}$ which increases
with the slab size. 
This behavior is suggestive of a Landau-Zener type of dielectric 
breakdown, as found in previous DMFT studies and in agreement with equilibrium
calculations of the tunneling amplitude for a quasiparticle thorugh
an insulating slab.

\section*{Acknowledgements}
We thank M.Sandri, M.Schir\`o
and J.Han for insightful discussions. A.A. and M.C. are financed by the European Union under FP7 ERC Starting
Grant No. 240524 ``SUPERBAD''.  Part of this work was supported
by European Union, Seventh Framework Programme FP7, under Grant No. 280555 ``GO FAST''.

\appendix
\section{Details on the variational dynamics}
\label{app:eqs_of_motion}

A straightforward differentiation of Eq.~(\ref{eq:Hstar}) with respect to the variational matrices $\hat{\Phi}_i$ leads to the equation of motions
\begin{equation}
  \begin{split}
    &i \frac{\partial}{\partial t} 
    \left(
      \begin{array}{c}
        \Phi_{z,0}(t)\\
        \Phi_{z,1}(t)\\
        \Phi_{z,2}(t)
      \end{array}
    \right)
    =\\
    &\left(
      \begin{array}{ccc}
        h_{00}(z,t) & h_{01}(z,t) & 0\\
        h_{01}^*(z,t) & 0 & h_{01}(z,t)\\
        0 & h_{01}^*(z,t) & h_{22}(z,t)
      \end{array}
    \right)
    \left(
      \begin{array}{c}
        \Phi_{z,0}(t)\\
        \Phi_{z,1}(t)\\
        \Phi_{z,2}(t)
      \end{array}
    \right)
  \end{split}
  \label{eq:dot_phi_matrices}
\end{equation}
with
\begin{equation}
  \begin{split}
    h_{00}(z,t) =& \frac{U}{2} - E_{z} +  \frac{\delta_z}{1-\delta_z^2} \,\Bigg\{
    2 \left|R_z \right|^2 \varepsilon_z \\
    & - \bigg[R^*_{z+1}\, R_z \, \Delta_z\; \big(1-\delta_{z,N}\big) + c.c.\bigg] \\
    & - \bigg[R^*_{z-1}\,R_z \, \Delta^*_{z-1}\;\big(1-\delta_{z,1})+ c.c.\bigg] \\
     &+ \bigg[\delta_{z,1} \, R^*_1 \, \Gamma_{L} + \delta_{z,N}\, R^*_N \, \Gamma_{R} + c.c. 
     \bigg]\Bigg\},
  \end{split}
  \label{eq:h00}
\end{equation}

\begin{equation}
  \begin{split}
    h_{22}(z,t) =& \frac{U}{2} + E_{z} -  \frac{\delta_z}{1-\delta_z^2} \,\Bigg\{
    2 \left|R_z \right|^2\,  \varepsilon_z \\
     & - \bigg[ R^*_{z+1}\, R_z \, \Delta_z \, \big(1-\delta_{z,N})+ c.c.\big) \bigg] \\
    & - \bigg[R^*_{z-1}\, R_z \, \Delta^*_{z-1}\,\big(1-\delta_{z,1}\big)+ c.c.\bigg]\\
     &+ \bigg[\delta_{z,1} R^*_1 \Gamma_{L} + \delta_{z,N} R^*_N \Gamma_{R} + c.c. \bigg]
     \Bigg\},
  \end{split}
  \label{eq:h22}
\end{equation}

\begin{equation}
  \begin{split}
    h_{01}(z,t) =&  \frac{\sqrt{2}}{\sqrt{1-\delta_z^2}} \bigg[
    R^*_z\, \varepsilon_z - R^*_{z+1}\, \Delta_z\,\big(1-\delta_{z,N}\big)\\
    &\!\!\!\!\!\! \!\!\!\!\! \!\!\!\!\!\! \!\!\!\!\!
     - R^*_{z-1}\,\Delta^*_{z-1}\,\big(1-\delta_{z,1}\big) + \delta_{z,1}\, \Gamma^*_{L} + 
    \delta_{z,N}\, \Gamma^*_{R}\bigg].
  \end{split}
  \label{eq:h01}
\end{equation}
The quantities appearing in the equations of motion (\ref{eq:h00}-\ref{eq:h01})
are defined by quantum averages of fermionic operators over the uncorrelated wavefunction $\ket{\Psi_0(t)}$
\begin{equation}
  \begin{split}
    &\varepsilon_z(t) = \sum_{\bk \sigma}\, \braket{\Psi_0(t)}{\, \dc_{\bk z \sigma } \da_{\bk z \sigma}\, }{\Psi_0(t)}\\
    &\Delta_z(t) = \sum_{\bk \sigma}\, \braket{\Psi_0(t)}{\, \dc_{\bk z+1 \sigma } \da_{\bk z \sigma}\, }{\Psi_0(t)}\\
    &\Gamma_{\alpha}(t) = \sum_{\bk \sigma}\,  \sum_{\kp} v_{\kp} \braket{\Psi_0(t)}{\, \dc_{\bk z_{\alpha} \sigma } \ca_{\bk \kp \alpha \sigma}\, }{\Psi_0(t)},
  \end{split}
  \label{eq:psi0_ave}
\end{equation}
and their time evolution is determined by the effective Scr\"odinger equation (\ref{eq:dot_psi}).

To solve for the dynamics of the effective Hamiltonian we introduce the Keldysh Greens' functions on the uncorrelated
wavefunction for $c$ and $d$ operators
\begin{eqnarray}
  \mathcal{G}^{K}_{\bk \sigma}(z,z';t,t') &=& -i\, 
  \quave{T_K\, \bigg(\da_{\bk z \sigma}(t) \, \dc_{\bk z' \sigma}(t')\bigg)}\\
  \label{eq:Gslab}
  g^{K}_{\bk \kp \alpha \sigma}(z;t,t') &=& -i\, \quave{T_K\,  \bigg(\ca_{\bk \kp \alpha \sigma}(t) \dc_{\bk z' \sigma}(t')\bigg)}
  \label{eq:Ghyb}
\end{eqnarray}
and express the quantities in Eqs.~\ref{eq:psi0_ave} in terms of their lesser components computed at equal time
\begin{equation}
  \begin{split}
    \quave{\dc_{\bk z \sigma } \da_{\bk z' \sigma}} (t) =& -i\,  \mathcal{G}^<_{\bk \sigma}(z',z;t,t) \\
    \quave{\dc_{\bk z \sigma } \ca_{\bk \kp \alpha \sigma}} (t) =& -i \, g^<_{\bk \kp \alpha \sigma}(z ;t,t).
  \end{split}
  \label{eq:Gless}
\end{equation}
We compute the equations of motion for the lesser components at equal times, Eq.~(\ref{eq:Gless}),
using the Heisenberg evolution for operators $c$ and $d$ with Hamiltonian  $\mathcal{H}_*$.
In order to get a closed set of differential equations we have to further introduce
the dynamics for the leads lesser Green function, which due to the hybridization
with the slab lose its translational invariance in the $z$-direction
\begin{equation}
  \bigg[G^{\alpha \alpha'}_{\bk \kp \kp' \sigma}\bigg]^{<}(t,t) = i\quave{\cc_{\bk \kp \alpha \sigma} \ca_{\bk \kp' \alpha' \sigma}}.
\end{equation}
Dropping, for the sake of simplicity, the lesser symbol and the spin index we get for each $\bk$ point the following equations of motion
\begin{widetext}
\begin{equation}
\begin{split}
    i \partial_t
    \mathcal{G}_{\bk}(z,z') =& \;\epsilon_{\bk}\Big( |R_z|^2 - |R_{z'}|^2 \Big)\, \mathcal{G}_{\bk}(z,z') 
    + \sum_{i=\pm1} \, R_{z+i}^*\,R_z \,\mathcal{G}_{\bk}(z+i,z') - R_{z}^*\,R_{z+i}\, \mathcal{G}_{\bk}(z,z'+i)  \\
    & + \sum_{\alpha=L,R}\, \delta_{z,z_{\alpha}} \,R^*_{z_{\alpha}} \,\sum_{\kp} \,v_{\kp}^{\alpha} \, g_{\bk \kp}^{\alpha}(z') \, + 
     \sum_{\alpha=L,R}\, \delta_{z' z_{\alpha}}\, R_{z_{\alpha}} \,\sum_{\kp}\, v_{\kp}^{\alpha} \Big[g_{\bk \kp}^{\alpha}(z)\Big]^*,
  \end{split}
  \label{eq:dot_Gslab}
\end{equation}

\begin{equation}
  \begin{split}
    i\partial_t g_{\bk \kp}^{\alpha}(z) =&\;  \Big(\varepsilon_{\bk}^{\alpha} + t_{\kp}^{\alpha} \Big)
    \, g_{\bk \kp}^{\alpha}(z) 
    - R^*_{z+1}\, R_z \; g_{\bk \kp}^{\alpha}(z+1) - R^*_{z-1}\,R_z \; g_{\bk \kp}^{\alpha}(z-1) + 
    v_{\kp}^{\alpha} \, R_{z_{\alpha}}\,  \mathcal{G}_{\bk} (z_{\alpha},z) \\
    & - \sum_{\alpha'=L,R} \, \delta_{z_{\alpha'},z}\,  \sum_{\kp}\,  v_{\kp}^{\alpha}\, R_{z_{\alpha'}} 
    \, G_{\bk \kp \kp'}^{\alpha \alpha'},
  \end{split}
  \label{eq:dot_Ghyb}
\end{equation}

\begin{equation}
  \begin{split}
    i\partial_t G_{\bk \kp \kp'}^{\alpha \alpha'} =& \Big( t_{\kp}^{\alpha} - t_{\kp'}^{\alpha'} \Big) \, 
    G_{\bk \kp \kp'}^{\alpha \alpha'} 
     - v_{\kp'}^{\alpha'}\, R^*_{z_{\alpha'}} \; g_{\bk \kp}^{\alpha}(z) - v_{\kp}^{\alpha} \, R_{z_{\alpha'}}\; \Big[g_{\bk \kp'}^{\alpha'}(z)\Big]^*.
  \end{split}
  \label{eq:dot_Gleads}
\end{equation}
\end{widetext}
The set of differential equations, composed by 
Eqs.~(\ref{eq:dot_Gslab}-\ref{eq:dot_Gleads}) and \ref{eq:dot_phi_matrices},
completely determines the dynamics within the time dependendent Gutzwiller 
and it is solved using a standard 4$-th$ order implicit Runge-Kutta method.\cite{brunner}
We mention that this strategy for the solution of the Gutzwiller dynamics
correspond to a discretization of the semi-infinite metallic leads. In principle,
the latter can be integrated-out exactly at the cost of solving the dynamics
for the lesser($<$) and greater($>$) component of the Keldysh Greens' function on the
whole two times $(t,t')$-plane.
However, such a route can be extremly costly from a computational point of
view and restric the simulations to small evolution times.
We explicitly checked that the dynamics using the above leads discretization coincides
with the dynamics obtained with the two time  $(t,t')$-plane evolution,
up to times for which finite size effects occour.
The latter can be however pushed far away with respect to the maximum times reachable
within the two time  $(t,t')$-plane evolution.

\section{Landau-Zener stationary tunneling within the Gutzwiller approximation}
\label{app:Landau-Zener}

We believe it is instructive to explicitly show how the Landau-Zener stationary tunnelling across the Mott-Hubbard gap in the presence of a voltage drop  translates into the language of the TDG approximation. 
Here, the gap and the voltage bias are actually absorbed into layer-dependent hopping renormalization factors $R_z(t)$ so that, an electron entering the Mott insulating slab from the metal lead translates into a free quasiparticle with hopping parameters that decay exponentially inside the insulator. In other words, quasiparticles within the Gutzwiller approximation do not experience a tunneling barrier in the insulating side 
but rather an exponentially growing mass. 

From this viewpoint, the {\sl living layer} that appears at the metal-Mott insulator interface can be legitimately regarded as the evanescent wave yielded by tunnelling across the Mott-Hubbard gap.  
Such a correspondence can be made more explicit following
Ref.~\onlinecite{Borghi_prb10} and its Supplemental Material.

Specifically, we shall consider a single metal-Mott insulator interface at equilibrium, with the metal and the Mott insulator confined in the regions $z<0$ and $z\geq 0$, respectively. 
The new ingredient that we add with respect to Ref.~\onlinecite{Borghi_prb10} is an electrochemical potential $\mu(z)$, 
which is constant and for convenience zero on the metal side, i.e. 
$\mu(z<0)=0$, while finite on the insulating side, $\mu(z\geq 0)\not = 0$, thus mimicking the bending of the Mott-Hubbard side bands at the junction. 

If the correlation length $\xi$ of the Mott insulator is much bigger that the inverse Fermi wavelength, 
in the Gutzwiller approach we can further neglect as a first approximation the $z$-dependence of the averages of hopping operators over the uncorrelated Slater determinant 
$\mid\!\Psi_0\rangle$.~\cite{Borghi_prb10}  
We can thus write the energy of the system as a functional of the variational matrices only, 
\begin{widetext}
\bea
E &=& -\frac{2}{24\,L}\,\sum_z\,R(z)^2 - \frac{1}{24\,L}\sum_z\, R(z)\,R(z+1) 
 +  \fract{1}{2\,L}\,\sum_z\, U(z)\,\Big( \mid\!\Phi_0(z)\!\mid^2 + \mid\!\Phi_2(z)\!\mid^2 \Big)
- \frac{1}{L}\,\sum_z\,\mu(z)\,\delta(z),\;\; \label{app:E}
\eea
\end{widetext}
where 
\ba
R(z) &=& \sqrt{\fract{2}{1-\delta(z)^2}}\;\;\Big(\Phi_1(z)^*\,\Phi_0(z)^\dagga + \Phi_2(z)^*\,\Phi_1(z)^\dagga\Big),
\ea
is the hopping renormalization factor, and 
\[
\delta(z) = \mid\Phi_0(z)\mid^2 - \mid\Phi_2(z)\mid^2,
\]
is the doping of layer $z$ with respect to half-filling, i.e. $n(z)=1-\delta(z)$. We have chosen units such that the Mott transition occurs at $U=1$, so that $U(z<0)=U_\text{metal}\ll 1$ on the metal side, and $U(z\geq 0)=U\gtrsim 1$ on the insulating one.

The minimum of $E$ in Eq. \eqn{app:E} can be always found with 
real parameters $\Phi_n(z)$, so that, since 
\[
 \Phi_0(z)^2 + \Phi_1(z)^2  + \Phi_2(z)^2 =1, 
 \]
there are actually two independent variables per layer. We can always choose these variables as $R(z)\in[0,1]$ and 
$\delta(z)\in[-1,1]$, in which case 
\ba
\mid\Phi_0(z)\mid^2 + \mid\Phi_2(z)\mid^2 &=& \fract{1}{2}\,\Bigg(
\Xi\big[R(z),\delta(z)\big] \\
&& \qquad\qquad + \fract{\delta(z)^2}{\Xi\big[R(z),\delta(z)\big]}\Bigg),
\ea
where 
\ba
\Xi\big[R(z),\delta(z)\big] &=& 1 - \sqrt{1-R(z)^2}\;\;\sqrt{1-\delta(z)^2}\\
&&\simeq 1 - \sqrt{1-R(z)^2} + \fract{\delta(z)^2}{2}\, \sqrt{1-R(z)^2},
\ea
the last expression being valid for small doping. Minimizing $E$ in Eq.~\eqn{app:E} with respect to 
$\delta(z)$ leads to 
\bea
\delta(z) &\simeq& \fract{4\mu(z)}{U}\; \fract{1-\sqrt{1-R(z)^2}}{1+R(z)^2+\sqrt{1-R(z)^2}},
\label{app:delta}
\eea
for $z\geq 0$, and $\delta(z)=0$ for $z<0$.

Through Eq.~\eqn{app:delta} we find an equation for $R(z)$ 
in the insulating side $z\geq 0$ that, after taking the continuum limit, reads
\be
\fract{\partial^2 R(z)}{\partial z^2} = - \fract{\partial}{\partial R(z)}\;V\Big[R(z),z\Big],
\ee 
which looks like a classical equation of motion with $z$ playing the role of time $t$, $R(z)$ that of 
the coordinate $q(t)$, and $V$ that of a time-dependent potential 
\bea
V\big(q,t\big) &=& -6U\,\bigg(1-\sqrt{1-q^2}\bigg) + 3q^2\nonu \\
&& + \fract{48\mu(t)^2}{U}\;\fract{1-\sqrt{1-q^2}}{1+q^2+\sqrt{1-q^2}}.
\eea
On the metallic side $R(z<0)\simeq R_\text{metal}\simeq 1$, so that the role of the junction is translated into appropriate boundary conditions at $z=0$.
 
Far inside the insulator, $R(z)\ll 1$ and we 
can expand 
\ba
V\Big[R(z),z\Big] &\simeq& \left(-3U+ 3 + 24\,\fract{\mu(z)^2}{U}\right)\,R(z)^2,
\ea
so that the linearized equation reads
\be
\fract{\partial^2 R(z)}{\partial z^2} = \left[6U - 6 - \fract{48}{U}\;\mu(z)^2\right]\;R(z),\label{app:shroedinger1}
\ee 
for $z>0$, while, in the metal side, $z<0$, where $R(z)$ is 
approximately constant, 
\be
\fract{\partial^2 R(z)}{\partial z^2} = 0.\label{app:shroedinger2}
\ee 
Equations \eqn{app:shroedinger1} and \eqn{app:shroedinger2} can be regarded as the Shr{\oe}dinger equation of a zero-energy particle 
impinging on a potential barrier at $z\geq 0$. Within the WKB approximation, the transmitted wavefunction at $z$ reads
\bea
R(z) \propto \exp\bigg(-\int_0^{z_*}\,d\zeta\,
\sqrt{6U - 6 - \fract{48}{U}\;\mu(\zeta)^2}\;\bigg),\;\;\;\label{app:transmitted}
\eea
where, assuming a monotonous $\mu(\zeta)$, the upper limit of integration is $z_*=z$ if 
$8\mu(z)^2 \leq U\,(U-1)$ otherwise is the turning point, i.e. 
$z_*$ such that $8\mu(z_*)^2 = U\,(U-1)$. 
 
Let us for instance take $\mu(z)=\text{E}\,z$, which corresponds to a constant electric field. In this case 
\be
|\text{E}|\,z_* = \sqrt{\fract{U(U-1)}{8}},
\ee
so that the transmission probability 
\be
\left|R(z>z_*)\right|^2 \sim 
\exp\Bigg(-\fract{E_\text{th}}{E}\Bigg),\label{app:R^2}
\ee
where the threshold field
\be
E_\text{th} = \fract{\pi}{2}\;\sqrt{\fract{U}{48}}\;\xi^{-2},
\label{app:E_th}
\ee
with the definition of the correlation length $\xi^{-1}=\sqrt{6(U-1)}$ of Ref.~\onlinecite{Borghi_prb10}. 

We observe that Eq.~\eqn{app:R^2} has exactly the form predicted by the Zener tunnelling in a semiconductor upon identifying 
\be
E_g\,\sqrt{\fract{m_*\,E_g}{\hbar^2}} \sim U-U_c,
\ee
where $E_g$ is the semiconductor gap, $m_*$ the mass parameter and 
$U_c$ the dimensional value of the interaction at the Mott transition.

\subsection{Growth of the living layer}
\label{Growth of the living layer}

The same approximate approach just outlined can be also extended 
away from equilibrium. We shall here consider the simple case of constant and vanishing electrochemical potential $\mu(z)=0$. 
We need to find the saddle point of the action
\be
S = \int dt\, i\sum_{n=0}^2\sum_{z}\,\Phi_{n}(z,t)^* \, 
\dot{\Phi}_n(z,t)^\dagga  - E(t),\label{app:S}
\ee
where $E(t)$ is the same functional of Eq. \eqn{app:E} where now all 
parameters $\Phi_n(z,t)$ are also time dependent. At $\mu(z)=0$ we can set 
\bea
\Phi_0(z,t) &=& \Phi_2(z,t) = \fract{1}{\sqrt{2}}\;
\text{e}^{i\phi(z,t)}\;\sin\fract{\theta(z,t)}{2},\label{app:Phi0(t)}\\
\Phi_1(t) &=& \cos\fract{\theta(z,t)}{2}, \label{app:Phi1(t)}
\eea
so that the equations of motion read
\bea
\sin\theta(z,t)\,\dot{\phi}(z,t) &=& -2\,\fract{\partial E}{\partial \theta(z,t)},\label{app:dphi}\\
\sin\theta(z,t)\,\dot{\theta}(z,t) &=& 2\,\fract{\partial E}{\partial \phi(z,t)}.\label{app:dtheta}
\eea
Upon introducing the parameters
\bea
\sigma_x(z,t) &=& \sin\theta(z,t)\,\cos\phi(z,t),\label{app:sigma_x}\\
\sigma_y(z,t) &=& \sin\theta(z,t)\,\sin\phi(z,t),\label{app:sigma_y}\\
\sigma_z(z,t) &=& \cos\theta(z,t),\label{app:sigma_z}
\eea
where $\sigma_x(z,t)=R(z,t)$ is the time dependent hopping renormalizaton, the equations of motion can be written as 
\bea
\dot{\sigma}_x(z,t) &=& -2\sigma_y(z,t)\,
\fract{\partial E}{\partial \sigma_z(z,t)} = \frac{U}{2}\,
\sigma_y(z,t),\label{app:dsigma_x}\\
\dot{\sigma}_y(z,t) &=& 2\sigma_x(z,t)\,
\fract{\partial E}{\partial \sigma_z(z,t)}
-2\sigma_z(z,t)\,
\fract{\partial E}{\partial \sigma_x(z,t)} \nonumber\\
&& = -\frac{U}{2}\,
\sigma_x(z,t) -2\sigma_z(z,t)\,
\fract{\partial E}{\partial \sigma_x(z,t)}
,\label{app:dsigma_y}\\
\dot{\sigma}_z(z,t) &=& 2\sigma_y(z,t)\,
\fract{\partial E}{\partial \sigma_x(z,t)},\label{app:dsigma_z}
\eea
where 
\bea
\fract{\partial E}{\partial \sigma_x(z,t)} 
&=& -\frac{1}{6}\,\sigma_x(z,t) 
-\frac{1}{24}\,\Big(\sigma_x(z+1,t)+\sigma_x(z-1,t)\Big)\nonumber\\
&&\simeq -\frac{1}{4}\,\sigma_x(z,t) 
-\frac{1}{24}\,\fract{\partial^2\sigma_x(z,t)}{\partial z^2}.
\label{app:dE/dsigma}
\eea
The Eqs. \eqn{app:dsigma_x}--\eqn{app:dsigma_z} show that the Gutzwiller equations of motion actually coincide to those of a 
Ising model in a transverse field treated within 
mean-field, as originally observed in Ref.~\onlinecite{SchiroFabrizio_short}.

Inside the Mott insulating slab we can safely assume $\sigma_z(z,t)\sim 1$ and obtain the equation for $R(z,t)=\sigma_x(z,t)$
\be
\ddot{R}(z,t) = -\frac{U}{4}\,\big(U-1\big)\,R(z,t) 
+\fract{U}{24}\,\fract{\partial^2 R(z,t)}{\partial z^2},
\label{app:ddotR}
\ee
which is the time dependent version of Eq. \eqn{app:shroedinger1} 
and is just a Klein-Gordon equation 
\be
\fract{1}{c^2}\,\ddot{R} - \nabla^2\,R + m^2\,c^2\,R = 0,
\label{app:KG}
\ee
with light velocity $c$ and 
mass $m$ given by
\bea
c^2 &=& U/24,\\
m^2\,c^2 &=& 6\,\big(U-1\big) = \xi^{-2}.
\eea
In dimensionless units 
\ba
\fract{z}{\xi} \rightarrow z,\qquad 
\fract{c t}{\xi} \rightarrow t.
\ea
Eq. \eqn{app:KG} reads
\be
\ddot{R} - \nabla^2\,R + R = 0.
\label{app:KG-2}
\ee

Let us simulate the growth of the "living layer" by a single metal-Mott insulator interface and absorb the 
role of the metal into an appropriate boundary condition for the surface $z=0$ of the Mott insulator side $z\geq 0$. Specifically, we shall assume that initially $R(z,0)=R_0(z)$, with $R_0(0) = R_0 >0$ and 
$R_0(z\to\infty)=0$, as well as that, at any time $t$, the value of $R(z,t)$ at the surface remains constant, i.e. 
$R(0,t) = R_0$, $\forall t$. We denote as $R_*(z)$ the 
stationary solution of Eq. \eqn{app:KG-2} with the boundary condition $R_*(0)=R_0$, that is 
\be
R_*(z) = R_0\,\text{e}^{-z}.\label{app:R_*}
\ee
One can readily obtain a solution of Eq. \eqn{app:KG-2} satisfying all boundary condition, which, 
after defining 
\be
\phi(x)=R_0(x)-R_*(x), \label{app:phi(x)}
\ee
reads 
\begin{widetext}
\bea
R(z,t) &=& R_*(z)  + \fract{\phi(z+t)+\theta(z-t)\,\phi(z-t)-\theta(t-z)\,\phi(t-z)}{2}\label{app:R(z,t)}\\
&& - \frac{t}{2}\, \int_{-t}^{t}\,dx\;\fract{J_1
\Big(\sqrt{t^2-x^2}\,\Big)}{\sqrt{t^2-x^2}}\Big[ \theta(x+z)\,\phi(x+z)-\theta(x-z)\,\phi(x-z)\Big]
,\nonumber 
\eea
\end{widetext}
where $J_1(x)$ is the first order Bessel function. We observe that for very long times 
$R(z,t\to\infty)\to R_*(z)$, namely the solution evolves into a steady state that corresponds to the 
equilibrium evanescent wave with the appropriate boundary condition. Moreover, Eq. ~\eqn{app:R(z,t)} 
also shows a kind of light-cone effect compatible with the full evolution that takes into account also the dynamics of the Slater determinant, which we have neglected to get Eq. \eqn{app:KG-2}. 
In fact, the missing Slater determinant dynamics is the reason why the initial exponential growth is not 
captured by Eq.~\ref{app:R(z,t)}, which thence has to be rather regarded as an asymptotic description valid only at long time and distances.

Another possible boundary condition is to impose that $\partial_z R(z,t)$ remains constant at $z=0$, rather 
than its value. In this case, if 
\be
A = -\fract{\partial R(z,0)}{\partial z}_{z=0} =  -\fract{\partial R_0(z)}{\partial z}_{z=0}\; ,
\ee
then we must take $R_*(z)=A\,\text{e}^{-z}$ and still $\phi(x)=R_0(x)-R_*(x)$ so that 
the solution reads
\begin{widetext}
\bea
R(z,t) &=& R_*(z)  + \fract{\phi(z+t)+\theta(z-t)\,\phi(z-t)+\theta(t-z)\,\phi(t-z)}{2}\label{app:R(z,t)-2}\\
&& - \frac{t}{2}\, \int_{-t}^{t}\,dx\;\fract{J_1
\Big(\sqrt{t^2-x^2}\,\Big)}{\sqrt{t^2-x^2}}\Big[ \theta(x+z)\,\phi(x+z)+\theta(x-z)\,\phi(x-z)\Big]
.\nonumber 
\eea
\end{widetext}
Also in this case $R(z,t)$ evolves towards a stationary value that, in dimensional units, reads
\be
R(z,t\to\infty) = A\,\xi\,\text{e}^{-z/\xi},
\ee
hence growths exponentially at fixed $A$ and $z$ as the Mott transition is approached. 

\bibliographystyle{apsrev}
\bibliography{mybiblio}

\end{document}